\documentclass[twocolumn,superscriptaddress,amsmath,amssymb]{revtex4-1}

\usepackage{graphicx}
\usepackage{csquotes}
\usepackage{gensymb}
\usepackage{standalone}

\usepackage{physics}

\newcommand{\gradperp}{\nabla_\perp}
\newcommand{\gradpar}{\nabla_\parallel}
\newcommand{\scalardiv}[1]{\nabla \vdot \qty(#1)}

\begin{document}

\title{Comparing Two- and Three-Dimensional Models of Scrape-Off-Layer Turbulent Transport}

\author{T. E. G. Nicholas}
\email[]{thomas.nicholas@york.ac.uk}
\affiliation{
	York Plasma Institute, 
	Department of Physics, University of York, 
	Heslington, York YO105DD, UK
}
\affiliation{
	United Kingdom Atomic Energy Authority,
	Culham Centre for Fusion Energy,
	Culham Science Centre,
	Abingdon, OX14 3DB, UK 
}

\author{J. Omotani}
\affiliation{
    United Kingdom Atomic Energy Authority,
	Culham Centre for Fusion Energy,
	Culham Science Centre,
	Abingdon, OX14 3DB, UK 
}

\author{F. Riva}
\affiliation{
	United Kingdom Atomic Energy Authority,
	Culham Centre for Fusion Energy,
	Culham Science Centre,
	Abingdon, OX14 3DB, UK  
}

\author{F. Militello}
\affiliation{
	United Kingdom Atomic Energy Authority,
	Culham Centre for Fusion Energy,
	Culham Science Centre,
	Abingdon, OX14 3DB, UK 
}

\author{B. Dudson}
\affiliation{
	York Plasma Institute, 
	Department of Physics, University of York, 
	Heslington, York YO105DD, UK
}
\affiliation{
	United Kingdom Atomic Energy Authority,
	Culham Centre for Fusion Energy,
	Culham Science Centre,
	Abingdon, OX14 3DB, UK  
}

\date{\today}
\begin{abstract}
There exists a large body of previous work using reduced two-dimensional models of the SOL, which model fluctuations in the drift-plane but approximate parallel transport with effective loss terms.
Full size three-dimensional simulations of SOL turbulence in experimental geometries are now possible, but are far more computationally expensive than 2D models.
We therefore use a flux-tube geometry model of the scrape-off layer to compare the results of 2D simulations to 3D simulations with a similar setup, looking for systematic differences.
Overall there is good agreement in the basic radial profiles, probability distribution functions, and power spectra of fluctuations.
However, the average temperature is over-predicted in 2D relative to 3D, and we explain the difference in terms of the effect of geometrical simplifications of devices at low power.
Varying geometric parameters, we find that supersonic flow in the divertor leg, which occurs because our simulations do not include neutrals and so represent low-recycling conditions, means that the divertor leg length only has a weak effect on the output.
Finally, we examine the effect of altering the magnitude of source and sink terms in 2D, concluding that they cannot easily be used to recreate both the density and temperature profiles observed in 3D simultaneously.
\end{abstract}

\maketitle

\section{Introduction}

Management of the heat flux to the divertor targets is a major challenge for developing a full-power tokamak fusion reactor, and especially so for a commercial power plant.
The exhaust power is fixed for a given design - dictated by the reactor's rated power output and energy gain factor $Q$ - but the radial profile of heat deposition sets the total area over which that heat is distributed.
These radial profiles are in turn set by the level of cross-field transport in the Scrape-Off Layer (SOL), compared with the rate of parallel heat transport towards the targets.

Cross-field transport in the SOL cannot be modelled as purely advective or diffusive, and instead requires resolving and then averaging turbulent fluctuations and short-lived structures\cite{Krasheninnikov2001}.

Drift-fluid turbulence modelling is computationally expensive, and drastically more so in three dimensions due to the need to resolve fast parallel timescales.
Therefore many reduced two-dimensional drift-plane models have been employed by the community, including ESEL\cite{Garcia2005, Militello2012}, SOLT\cite{Russell2009,Myra2011}, STORM-2D\cite{Easy2014,Easy2016}, and TOKAM2D\cite{Bisai2005}.
These models evolve variables in the drift-plane, but simplify the parallel transport along the magnetic fields lines as overall loss terms.

It is now possible to perform full 3D scrape-off-layer turbulence simulations, which explicitly resolve fluctuations along the parallel direction.
Codes with 3D capability include GBS\cite{Halpern2016}, SOLT3D\cite{Umansky2019}, TOKAM3X\cite{Tamain2016}, GRILLIX\cite{Stegmeir2018}, and HERMES\cite{Dudson2017}.
Recently this was even demonstrated for full-scale simulations of the turbulent SOL in the medium-sized tokamak MAST\cite{Sykes2001}, using STORM-3D\cite{Riva2019}.
Several of these codes have been benchmarked against one another through simulations of the ISTTOK device\cite{Dudson2020}.

However, there exists a large catalogue of previous work done with the various 2D models (e.g. \cite{Militello2013,Garcia2006,Myra2011a,Bisai2005}, as presented in the comprehensive review by D'Ippolito\cite{DIppolito2011}), which successfully reproduce key experimental results such as the universality of fluctuation statistics\cite{Garcia2004}.
These codes continue to possess some advantages over the 3D models, such as computational speed, model simplicity, and physical interpretability.


We present a systematic comparison of 2D models with their closest 3D analogues, providing a basis to evaluate the systematic errors of the approximations used to truncate the models to 2D.

In section \ref{sect:model} we describe the systems of equations solved, before describing the parameters chosen and the domain geometry in section \ref{sect:numerics}.
Our results begin with a comparison of 2D and 3D simulations (section \ref{sect:Comparison2Dvs3D}), before studying the effect of varying the divertor leg length (section \ref{sect:leglength}), and the sensitivity to some of the overall simulation parameters (section \ref{sect:sensitivity}).

\section{Physical Model}
\label{sect:model}

The model and code we will use is STORM, a scrape-off layer physics module built using the BOUT++ framework for plasma fluid simulations\cite{Dudson2009,Dudson2014,Dudson2016a}.
STORM has two related models: a 3D version\cite{Riva2019,Walkden2019, Walkden2015}, and a 2D version\cite{Easy2014}.
Because of the modular way that BOUT++ is designed, these two versions of the STORM code use mostly the same methods to solve the perpendicular dynamics.

STORM has previously been used to model individual filaments\cite{Easy2014,Easy2016a}, finite electron temperature effects\cite{Walkden2016}, filament interactions\cite{Militello2017}, electromagnetic effects\cite{Hoare2019}, neutral background\cite{Schworer2017,Schworer2018}, divertor turbulence\cite{Walkden2019}, and full 3D geometries\cite{Riva2019}.

\subsection{Three-dimensional STORM model}

\newcommand{\curv}[1]{\mathcal{C}\qty(#1)}

We use a drift-reduced, fluid, cold-ion, electrostatic model, which evolves density $n$, $E \cross B$ vorticity $\Omega$, temperature $T$, electrostatic potential $\phi$, parallel electron velocity $V$, and parallel ion velocity $U$.
The equations are ultimately derived from the well-known Braginskii equations\cite{BraginskiiS1965}, where the Boussinesq approximation has been employed in deriving the vorticity equation \eqref{eq:STORMvorticity}.

\begin{multline}\label{eq:STORMdensity}
    \pdv{n}{t} = - \qty{\phi, n} - V \gradpar n - n \nabla_\parallel V + \curv{p} \\
    - n\curv{\phi} + D_n \laplacian_\perp{n} + S .
\end{multline}

\begin{multline}\label{eq:STORMionparallel}
    \pdv{U}{t} = -\qty{\phi, U} - U \gradpar U - \gradpar \phi \\
    - \frac{\nu_\parallel}{\mu} (U-V) +  0.71 \gradpar T - U\frac{S}{n}
\end{multline}

\begin{multline}\label{eq:STORMelectronparallel}
    \pdv{V}{t} = -\qty{\phi, V} - V \gradpar V - \mu \gradpar \phi - \frac{\mu}{n} \gradpar p \\
    + \nu_\parallel (U-V) -  0.71 \mu \gradpar T - V\frac{S}{n}
\end{multline}

\begin{multline}\label{eq:STORMvorticity}
    \pdv{\Omega}{t} = - \qty{\phi, \Omega} - U \gradpar \Omega \\
    + n \qty[\gradpar (U-V) + (U-V) \gradpar \log(n)] \\
    + \curv{p} + \mu_\Omega \laplacian_\perp{\Omega}
\end{multline}

\begin{multline}\label{eq:STORMtemperature}
    \pdv{T}{t} = -\qty{\phi, T} - \frac{2}{3} T \curv{\phi} - V \gradpar T - \frac{2}{3} T \gradpar V 
    \\ - \frac{2}{3n} \gradpar q_\parallel - \frac{2}{3} 0.71 (U-V) \gradpar T + \frac{2}{3n} \frac{\nu_\parallel}{\mu n} J^2_\parallel 
    \\ + \frac{5}{3} T \curv{T} + \frac{2T}{3n} \curv{p} + \frac{2}{3n} \kappa_{\perp,0} \laplacian_\perp{T}
    \\
    + \frac{2}{3n} S_E + \frac{2}{3n} \frac{1}{2 \mu} V^2 S - \frac{2}{3n} \frac{V^2}{2 \mu} \curv{p} - \frac{T}{n} S. 
\end{multline}

\begin{equation}\label{eq:STORMpotential}
    \Omega = \scalardiv{\frac{\gradperp \phi}{B^2}}
\end{equation}

Here we have used the quantities 
electron pressure $p = nT$,
parallel current $J_\parallel = n \qty(U - V)$,
and parallel heat flux $q_\parallel = - \kappa_{\parallel,0} T^{5/2} \gradpar T - 0.71 T J_\parallel $; 
as well as various operators such as
the total derivative $\dv{f}{t} = \pdv{f}{t} + \qty{\phi, f}$,
the Poisson bracket $\qty{\phi, f} = \vb{b} \cross \qty(\grad{\phi} \vdot \grad{f}) / B$,
parallel gradient $\gradpar{f} = \vb{b} \vdot \grad{f}$,
perpendicular gradient $\gradperp{f} = \grad{f} - \vb{b} \gradpar{f}$, and 
perpendicular Laplacian $\laplacian_\perp{f} = \scalardiv{\gradperp f}$.
$\mu$ is the electron to ion mass ratio $m_e/m_i$.

Also, within the simplified geometry described in section \ref{sect:numerics} the curvature operator acting on a scalar field $\curv{f}$ takes the form\cite{Easy2016}
\begin{equation}
    \curv{f} \equiv \curl(\frac{\vb{b}}{B}) \vdot \grad{f} \approx - \frac{2}{R_0 B_0} \pdv{f}{z}.
\end{equation}

A Bohm normalisation scheme has been used for these equations (defined for STORM in \cite{Easy2016} and \cite{Walkden2019}), in which lengths and times are normalised to the hybrid Larmor radius and the ion gyrofrequency
\begin{equation}
\begin{split}
    \frac{\vb{x}}{\rho_s} \rightarrow \vb{x}, \quad
    \Omega_i t \rightarrow t, \quad
    \frac{L_\parallel}{\rho_s} \rightarrow L_\parallel.
\end{split}
\end{equation}
Variables are normalised to characteristic values (for which the numerical values used are given in table \ref{tab:parameters}) or combinations of them.
The dimensionless forms of the evolved fluid variables are obtained from the dimensional forms through
\begin{equation}
\begin{split}
    n \equiv \frac{n_e}{n_{e,0}}, \quad
    T \equiv \frac{T_e}{T_{e,0}}, \quad
    \phi \equiv \frac{e \varphi}{T_{e,0}}, \\
    V \equiv \frac{u_{\parallel,e}}{c_s}, \quad
    U \equiv \frac{u_{\parallel,i}}{c_s}, \quad
    \Omega \equiv \frac{\omega}{\Omega_i},
\end{split}
\end{equation}
meaning currents and sources are normalised through
\begin{equation}
    J_\parallel \equiv \frac{j_\parallel}{e n_{e,0} c_s}, \quad
    S_n \equiv \frac{S^{\qty(n)}}{n_{e,0} \Omega_i}, \quad 
    S_E \equiv \frac{W}{n_{e,0} T_{e,0} \Omega_i},
\end{equation}
and the normalised parallel resistivity given by
\begin{equation}
    \nu_\parallel \equiv \frac{0.51 \nu_{ei}}{\Omega_i}.
\end{equation}
Diffusion coefficients are normalised to the Bohm diffusion ($D_\text{Bohm} \equiv \rho_s^2 \Omega_i$)
\begin{equation}
    D_n \equiv \frac{D}{D_\text{Bohm}}, \quad
    \mu_\Omega \equiv \frac{\mu_\omega}{D_\text{Bohm}}, \quad
    \kappa \equiv \frac{\kappa}{D_\text{Bohm}},
\end{equation}
and numerical values are given in table \ref{tab:diffusioncoefficients}.
For more details see \cite{Easy2016}.

The system \eqref{eq:STORMdensity}-\eqref{eq:STORMtemperature} requires a corresponding set of boundary conditions.
To model the interaction with the divertor target plates, we describe the plasma dynamics at the magnetic pre-sheath entrance by imposing Bohm sheath boundary conditions for the parallel velocities $U$ and $V$, and computing the corresponding parallel electron power flux $Q_\parallel$ by following \cite{Stangeby2000}.
Therefore at each target plate we require

\newcommand{\vfloat}{V_{\text{fl}}}

\begin{equation}\label{eq:sheathbcs}
\begin{split}
    U_\parallel |_\text{target} &\geq \sqrt{T} \\
    V_\parallel |_\text{target} &= 
    \begin{cases}
        \sqrt{T} \exp(-\vfloat -\frac{\phi}{T}),  & \text{if } \phi > 0, \\
        \sqrt{T} \exp(-\vfloat),          & \text{otherwise}
    \end{cases}
    \\
    Q_\parallel |_\text{target} &= \gamma n T V,
\end{split}
\end{equation}
with the signs reversed if $\mathbf{B}$ is directed away from the wall.
Here $\vfloat$ is the plasma floating potential, defined as
\begin{equation}
    \vfloat = \frac{1}{2} \log(\frac{2\pi}{\mu} \qty(1 + 1/\mu)),
\end{equation}
where $\gamma=2+\vfloat$ is the sheath transmission coefficient.
The parallel electron heat flux $q_\parallel$ is determined from the power flux by
\begin{equation}\label{eq:sheathbcsq}
    q_\parallel |_\text{target} = Q_\parallel |_\text{target} - \frac{5}{2} n T V - \frac{1}{2 \mu} n V^3.
\end{equation}
The binormal boundary is periodic, and the radial boundary conditions are described in appendix \ref{appendix:bcs}.

\subsection{Two-dimensional STORM model}\label{sect:closure2D}

The 2D model system is a simplification of the 3D system.
Grouping all terms with parallel dependence into loss terms, in the perpendicular drift-plane equations \eqref{eq:STORMdensity}, \eqref{eq:STORMvorticity}, \& \eqref{eq:STORMtemperature} become
\begin{equation}
\label{eq:STORMdensity2D}
    \pdv{n}{t} = \frac{1}{B} \qty{\phi, n} + \curv{p} - n \curv{\phi} + S_n + D_n \gradperp^2 n - n_\text{loss}
\end{equation}
\begin{equation}
\label{eq:STORMvorticity2D}
    \pdv{\Omega}{t} = \frac{1}{B} \qty{\phi, \Omega} + \frac{\curv{p}}{n} + \mu_\Omega \gradperp^2 \Omega - \Omega_\text{loss}
\end{equation}
\begin{multline}
\label{eq:STORMtemperature2D}
    \pdv{T}{t} = \frac{1}{B} \qty{\phi, T} - \frac{2}{3} T \curv{\phi} + \frac{2T}{3n} \curv{p} + \frac{5}{3} T \curv{T} 
    \\ + \frac{2}{3n} S_E - \frac{TS}{n} + \frac{2}{3n} \kappa_{\perp_0} \gradperp^2 T - T_\text{loss}
\end{multline}
with the Laplacian inversion for the electrostatic potential \eqref{eq:STORMpotential} unchanged.

We use the so-called sheath-dissipation closure\cite{Easy2016,Garcia2006}, in which density and temperature are assumed to be constant along the parallel direction, and we treat the parallel velocities as linearly varying between the Bohm sheath values, so that $\gradpar{U} \sim U / L_\parallel$.

This gives the loss rate of density and vorticity as
\begin{equation}
    n_\text{loss} = \frac{1}{L_\parallel} n V_\text{sh}\qty(\phi, T)
\end{equation}
\begin{equation}
    \Omega_\text{loss} = \frac{1}{L_\parallel} (V_\text{sh}\qty(\phi, T) - \sqrt{T}),
\end{equation}
where
\begin{equation}
    V_\text{sh}\qty(\phi, T) = V_{\text{sh}_0} \sqrt{T} e^{-\phi/T},
\end{equation}
and
\begin{equation}
    V_{\text{sh}_0} = e^{- \vfloat}.
\end{equation}
However, we linearize the exponential sheath factor around the floating potential (as is done for example in \cite{Easy2016}), giving
\begin{equation}
    V_\text{sh}\qty(\phi, T) = \sqrt{T} \qty[1 - \qty(\vfloat + \frac{\phi}{T})] .
\end{equation}

We choose the sheath-dissipation closure over the vorticity-advection closure (also described in \cite{Easy2016}) because we found that the vorticity advection closure was not adequate for turbulence simulations, providing too little damping at large scales, which lead to the domain filling with unphysically-large structures \cite{Easy2014}.

To model parallel loss of thermal energy we interpolate between two limits, following the approach of Myra\cite{Myra2011}.
The loss rates in the two regimes (sheath-limited and conduction-limited\cite{Stangeby2000}) are
\begin{align}
    q_{\parallel_\text{SL}} &= \qty(\gamma_e - \frac{3}{2}) n T V_\text{sh}\qty(\phi, T) \\
    q_{\parallel_\text{CL}} &= \frac{2}{7} \frac{1}{L_\parallel} \kappa_{\parallel,0} T^{7/2},
\end{align}
where we have again linearized the potential dependence of the sheath velocity.
These loss rates are incorporated into \eqref{eq:STORMtemperature2D} through
\begin{equation}
    T_\text{loss} = \frac{2}{3 n} \frac{1}{L_\parallel} q_\parallel .
\end{equation}

(As an aside, dropping Myra's $q_\text{FL}$ term for the flux-limited regime\cite{Myra2011} is equivalent to using Fundamenski's free-streaming expression\cite{Fundamenski2007} when the collisionality is high, as we will see it is in these simulations.)

Harmonically averaging these loss rates through
\begin{equation}
\label{eq:interpolation}
    \frac{1}{q_\parallel} = \frac{1}{q_{\parallel_\text{SL}}} + \frac{1}{q_{\parallel_\text{CL}}},
\end{equation}
effectively means that the loss is limited to the smaller of the two rates, with a smooth switchover controlled by the collisionality.
This ``heuristic'' parallel closure that we use is therefore equivalent to a linearized version of the ``sheath-dissipation'' closure, but which interpolates between two different regimes of parallel heat transport.

\section{Numerical Setup}
\label{sect:numerics}

The STORM code numerically solves \eqref{eq:STORMdensity}-\eqref{eq:STORMtemperature} in 3D, and \eqref{eq:STORMdensity2D} - \eqref{eq:STORMtemperature2D} in 2D.
The dissipative parameters ($\mu_n$, $\mu_\Omega$, $\kappa_\perp$, and $\kappa_\parallel$) were kept constant in time and space, and calculated using the classical expressions from \cite{Fundamenski2007}, evaluated using the reference normalisation values of density and temperature (given in table \ref{tab:parameters}).
These are the same values used to simulate the MAST SOL in \cite{Militello2012}, and are similar to those used for the full 3D MAST-U geometry STORM simulations performed in \cite{Riva2019}.
Ion temperature is set to zero in the derivation of the main model equations for simplicity, but is included in table \ref{tab:parameters} because a finite value is still required for some of the dissipative expressions from \cite{Fundamenski2007}.
The resistivity $\eta_\parallel$ was allowed to vary, following the $T^{-3/2}$ dependence that follows from the definition
\begin{equation}
    \eta_\parallel = 0.51 \frac{\nu_{ei_0}}{T^{3/2} \Omega_{e_0}}.
\end{equation}

\begin{table}[h!]
\centering
\begin{tabular}{ c c c c }
 \hline
 $n_0$ [$10^{-19}$m$^{-3}$] & $T_{e,0}$ [eV] & $T_{i,0}$ [eV] & $B_0$ [T] \\
 \hline
 $0.5$ & $15$ & $30$ & $0.24$ \\
 \hline
\end{tabular}
\caption{Reference normalisation values of density and temperature, used for all simulations.}
\label{tab:parameters}
\end{table}

\begin{table}[h!]
\centering
\begin{tabular}{ c c c c c c }
 \hline
 $D_n$ & $\mu_\Omega$ & $\kappa_\perp$ & $\kappa_\parallel$ & $\eta_\parallel \qty(T_e=T_{e,0})$ \\
 \hline
 $2.1 \vdot 10^{-4}$ & $1.6 \vdot 10^{-3}$ & $3.3 \vdot 10^{-4}$ & $4.4 \vdot 10^{4}$ & $3.6 \vdot 10^{-5}$ \\
 \hline
\end{tabular}
\caption{Normalised values of various dissipation coefficients used in all simulations.
Calculated using the standard classical expressions from the parameters given in table \ref{tab:parameters}.}
\label{tab:diffusioncoefficients}
\end{table}

\begin{figure}[ht]
\includegraphics[width=\columnwidth,trim={3.5cm 0.25cm 0.0cm 0},clip]{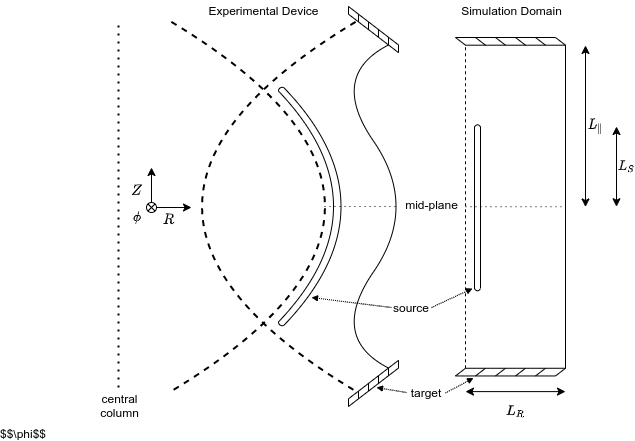}
\caption{Schematic diagram of the relationship between the simplified 3D simulation domain and the full experimental device geometry.
The 2D simulation domain here corresponds to a plane into the page located at the mid-plane.
The basis set of vectors $(R, Z, \phi)$ denote the radial, vertical and toroidal directions in the experimental domain, but are aligned along the radial, parallel and binormal directions in the simplified simulation domain.
In reality the simulation domain corresponds to a flux tube which is closer to horizontal than vertical (tilted into the page), because typically $B_\text{tor} > B_\text{pol}$ at the edge.
In this paper the simulation domain only models the region outside of the separatrix, marked by the dashed line.}
\label{fig:domainSOL}
\end{figure}

The equations were solved on a simplified cuboid domain (rectangular in 2D), which maps onto a schematic representation of the real SOL (figure \ref{fig:domainSOL}).
Vertically extending from target to target, the simplified domain can be interpreted as a straightened SOL flux tube, and has periodicity only in one direction.
The pitch of the field lines at the edge means that our periodic direction perpendicular to both the major radius and the magnetic field in our simulations is angled relative to the real toroidal direction in experiments, and so will hereon be referred to as the binormal direction instead.
The domain contains no magnetic shear, though this is a choice which could be relaxed in future work with STORM.

The numerical domain for the baseline 3D simulation spans $140.625 \rho_s$ in the radial ($x$) direction, $4000.0 \rho_s$ in the parallel ($y$) direction, and $150.0 \rho_s$ in the binormal ($z$) direction (i.e. $L_x=140.625$, $L_y=4000.0$ and $L_z=150$ in normalized units).
The 3D domain is resolved with $240 \times 32 \times 256 $ grid points, meaning that the grid cells were square in the perpendicular plane.
In 2D the domain covers the same ($x, z$) extent in normalized units, but a larger resolution is used of $960 \times 1024 $ grid points, and (by definition) only a single grid point in the $y$ direction.
Convergence tests were performed to assess the impact of the  perpendicular grid resolution, which showed that the grid used for 3D simulations is well resolved (see appendix \ref{appendix:resolutionscan}).

The simulations are source-driven, meaning that a volumetric density and energy source are present near the inner boundary of the domain (see figure \ref{fig:domainSOL}).
The source-driven simulations represent a SOL constantly fed with particles and power coming from the core plasma, and were chosen over fixing either the incoming flux or the pressure at the inner boundary so as to most directly compare the 2D and 3D models.
This choice was made to impose as few constraints as possible on the resultant variable and flux profiles, and separate the self-consistently generated profiles as much as possible from the inner boundary conditions.

The magnitude of the source terms was chosen so that the densities and temperatures in the 2D simulation were representative of the MAST SOL in L-mode near the ``separatrix'', i.e. around $1.0 \times 10^{-19}$m$^{-3}$ and $10$eV.
This represents a relatively low-power L-mode shot.
The sources in the 3D simulations were then set such that the total particles and energy injected per second was the same as in the 2D simulations.

The sources (both temperature and density) are constant in the binormal direction, but have a Gaussian profile in the radial direction, centered upon a point located at $ L_x/10$, and with a radial width of $L_x / 120$.
They are therefore extremely localized, and present only at one location on the inner side of the domain.
In 3D the sources also have parallel extent.
A top-hat function is used, so that the source extends from the mid-plane half-way to the target in both directions (so $L_S = 0.5 L_\parallel$ in figure \ref{fig:domainSOL}).
The point in the parallel direction at which the source ends is intended to roughly represent the position of X-point.
The baseline 3D simulation therefore represents an experimental regime in which the main source of particles and power is from the core, crossing over the separatrix between the two X-points.
The inner part of the domain is therefore not considered to be physical, meaning that profiles should be compared to one another only from the radial position of the sources outwards.

The sources create gradients which display growing instabilities, which eventually non-linearly couple to produce turbulence which drives transport in the radial direction.
Throughout the domain particles, momentum, and energy are lost - in 3D by fluxes through the targets at either end, and in 2D by heuristic loss terms approximating those same parallel processes.

A statistical steady-state was required to represent a saturated L-mode turbulence regime.
To obtain data from this state, each simulation was run with an initial ``spin-up'' phase which was then discarded, and only the following phase of statistical steady state was used for analysis.
The spin-up phase lasted typically around 5ms, while the saturated phase was typically of the order of several milliseconds, which corresponds to thousands of turbulence correlation times.

\begin{figure}[ht]
\includegraphics[width=\columnwidth,trim={0.1cm 0.25cm 0.0cm 0.2cm},clip]{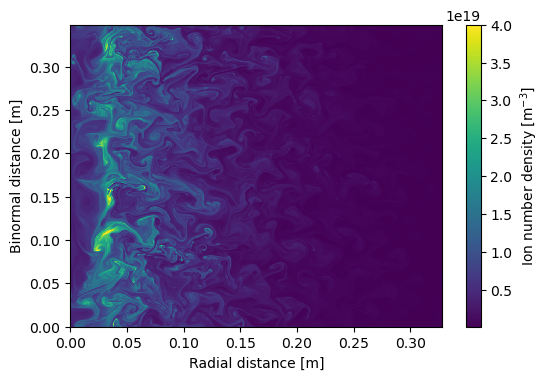}
\caption{Snapshot of the typical spatial variation of the density during the saturated phase of a simulation using the two-dimensional STORM model.}
\label{fig:typical_output}
\end{figure}

A typical output of the code, which shows a snapshot of the density fluctuations at the mid-plane of a 2D simulation, is shown in figure \ref{fig:typical_output}.

\section{Results and discussion}

\subsection{Comparison between 2D and 3D}
\label{sect:Comparison2Dvs3D}

There are multiple different metrics on which we can compare 2D and 3D simulations.
Each of these metrics will help us assess the suitability of the simplified 2D model for capturing some aspect of the more complex 3D physics, or for predicting some physical result of interest.
As a baseline we first compare a single 3D simulation with a single 2D simulation, set up so as to have the same sources and parallel connection length.

\subsubsection{Radial profiles}

The time-averaged radial profiles of quantities represent the balance achieved in steady-state between perpendicular transport and parallel loss.
In our model, which has hot electrons but cold ions, the fluid variables of interest that exist in both the 2D and 3D models are density, electron temperature, and potential.

\begin{figure}[ht]
\includegraphics[width=\columnwidth,trim={0.5cm 0.25cm 1cm 0},clip]{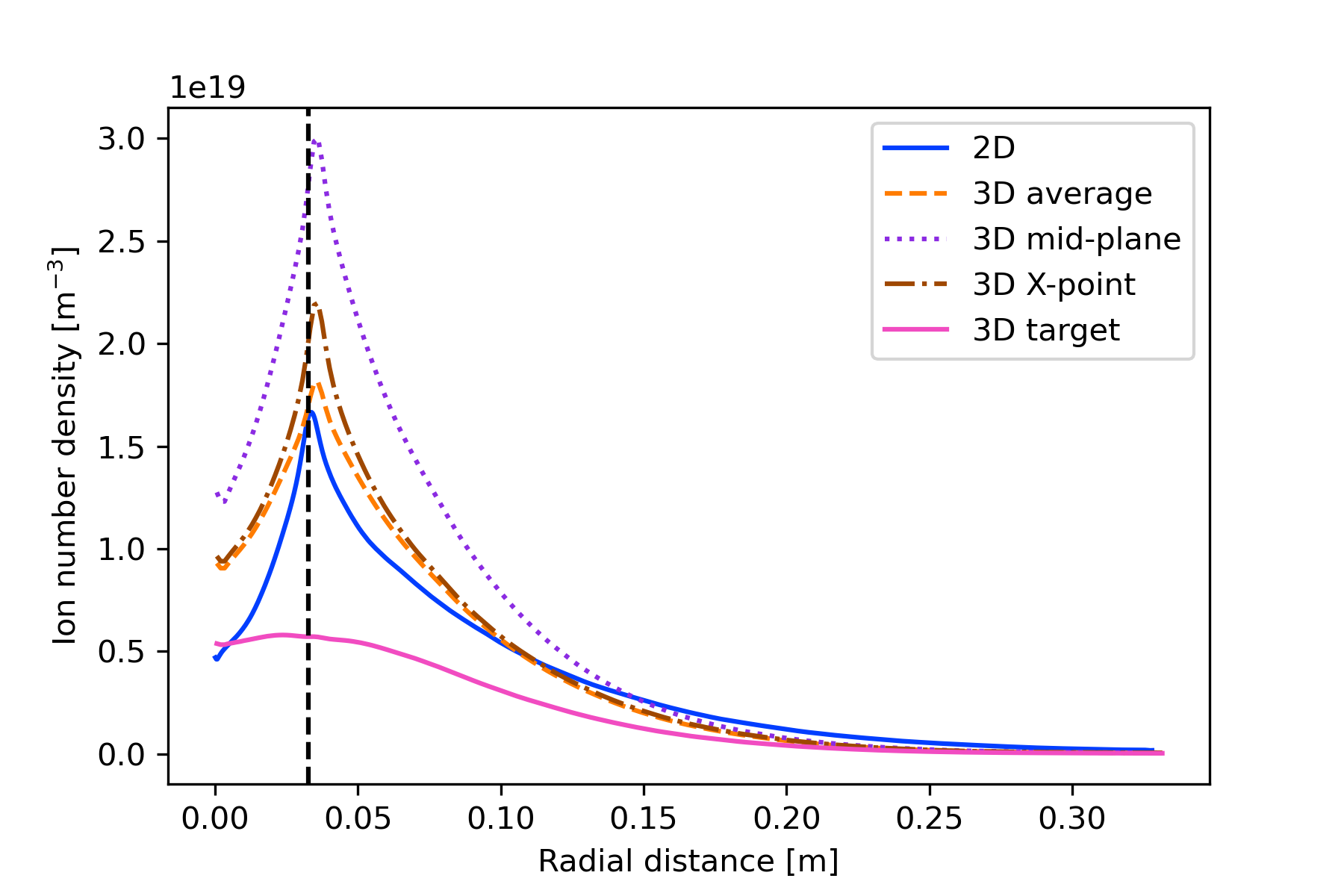}
\caption{Average radial profiles of density. Averages are performed over the saturated time periods, over the binormal direction, and in the case of the line labelled ``3D average'', also over the parallel domain.
The radial location of the density and temperature sources is shown by the black dotted line.}
\label{fig:nprofiles}
\end{figure}

\begin{figure}[ht]
\includegraphics[width=\columnwidth,trim={0.5cm 0.25cm 1cm 0},clip]{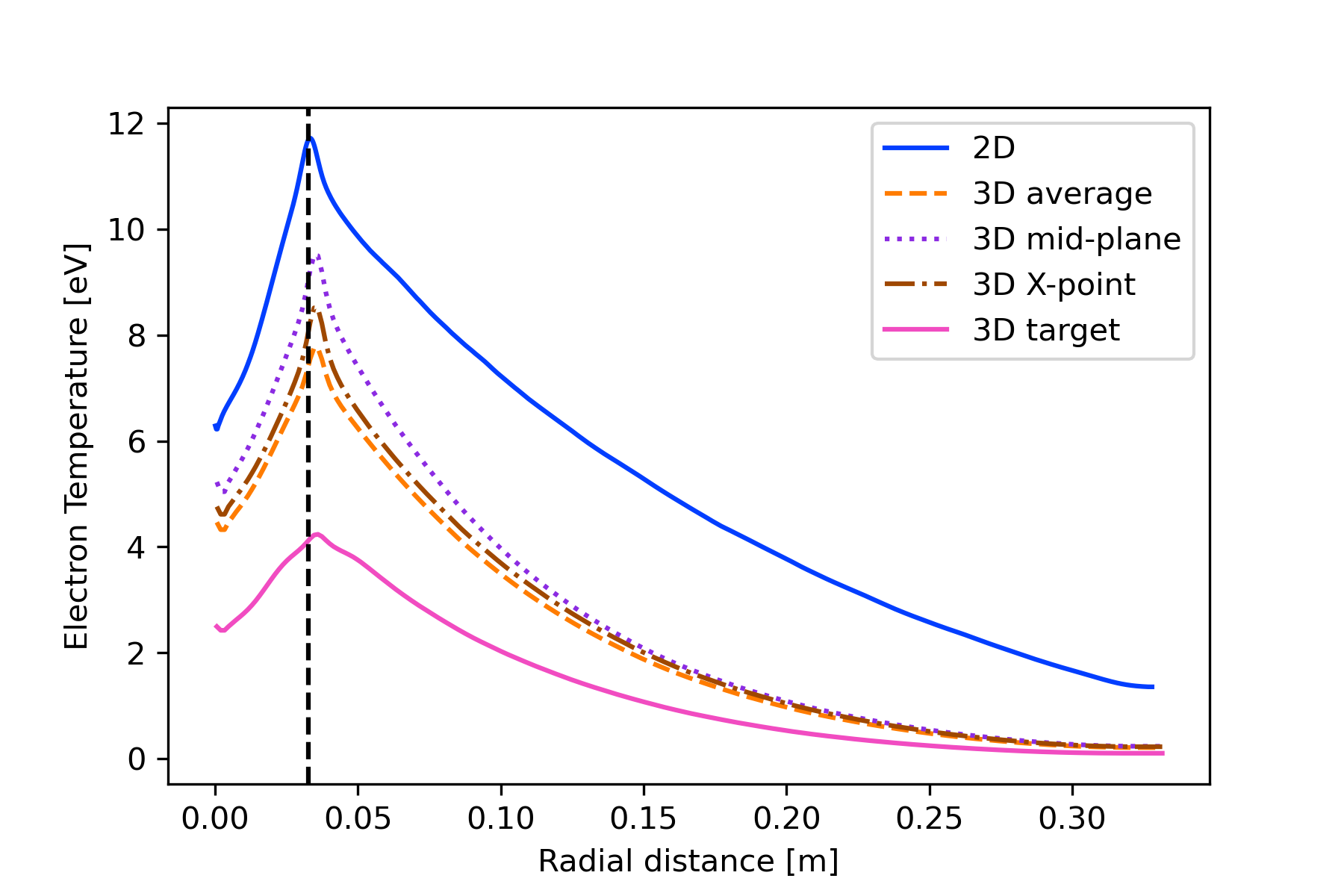}
\caption{Average radial profiles of electron temperature.
Averages are performed over the saturated time periods, over the binormal direction, and in the case of the line labelled ``3D average'', also over the parallel domain.
The radial location of the density and temperature sources is shown by the black dotted line.}
\label{fig:Tprofiles}
\end{figure}

We see in figure \ref{fig:nprofiles} that the density profile is well-captured by the 2D simulation: the exponential falloff length is reproduced (fitting a decaying exponential to $x>L_x/10$ gives $\lambda_n = 6.5$cm for the 2D line and $\lambda_n = 5.4$cm for the 3D average line), and the absolute value matches that of the parallel-averaged value of the 3D domain.
However, the choice of which 3D average we compare to matters - there is a factor of 2 difference between the mid-plane value or the parallel-averaged value.

The temperature is less well reproduced, being overestimated at all radial positions in 2D relative to 3D ($\lambda_T = 15.3$cm for the 2D line and $\lambda_T = 8.2$cm for the 3D average line).
This implies that the parallel heat loss is weaker (for the same upstream temperature) in 2D than in 3D, as discussed in section \ref{sect:heatloss}.
In the SOL the sheath causes characteristic potential fluctuations to be set by the temperature ($e \phi  \sim T$), so a similar difference is seen in the potential.

\subsubsection{Statistical properties}

Whilst the profiles tell us about the overall balance of fluxes, the statistical properties of the timeseries capture some of the dynamics of the turbulent fluctuations themselves.

\begin{figure}[ht]
\includegraphics[width=\columnwidth,trim={0.5cm 0.25cm 1cm 0},clip]{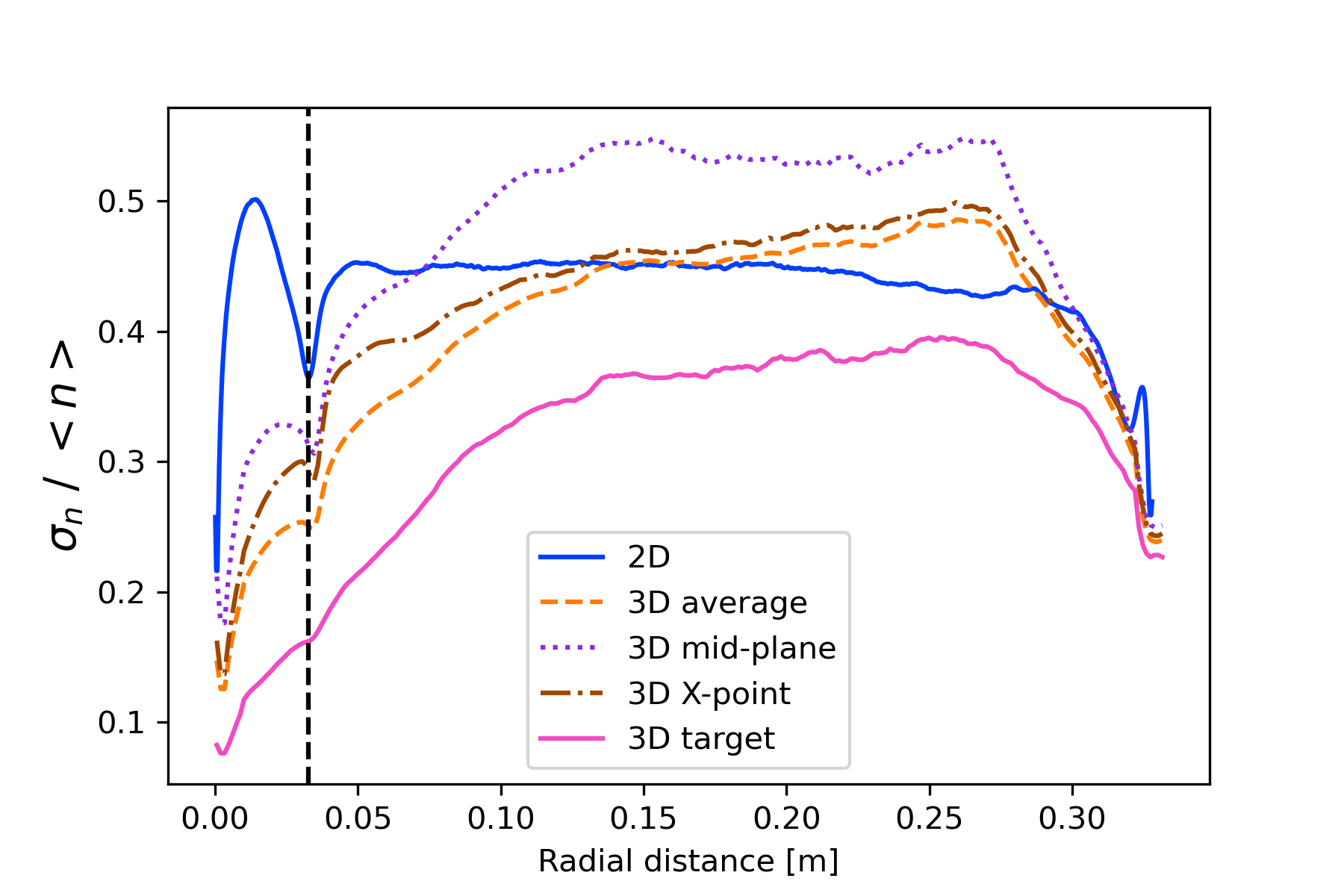}
\caption{Average radial profiles of the standard deviation of normalized density fluctuations away from the time-averaged mean local density.
Averages are performed over the saturated time periods, over the binormal direction, and in the case of the line labelled ``3D average'', also over the parallel domain.
The radial location of the density and temperature sources is shown by the black dotted line.}
\label{fig:nstddevprofiles}
\end{figure}

\begin{figure}[ht]
\includegraphics[width=\columnwidth,trim={0.2cm 0.25cm 1cm 0},clip]{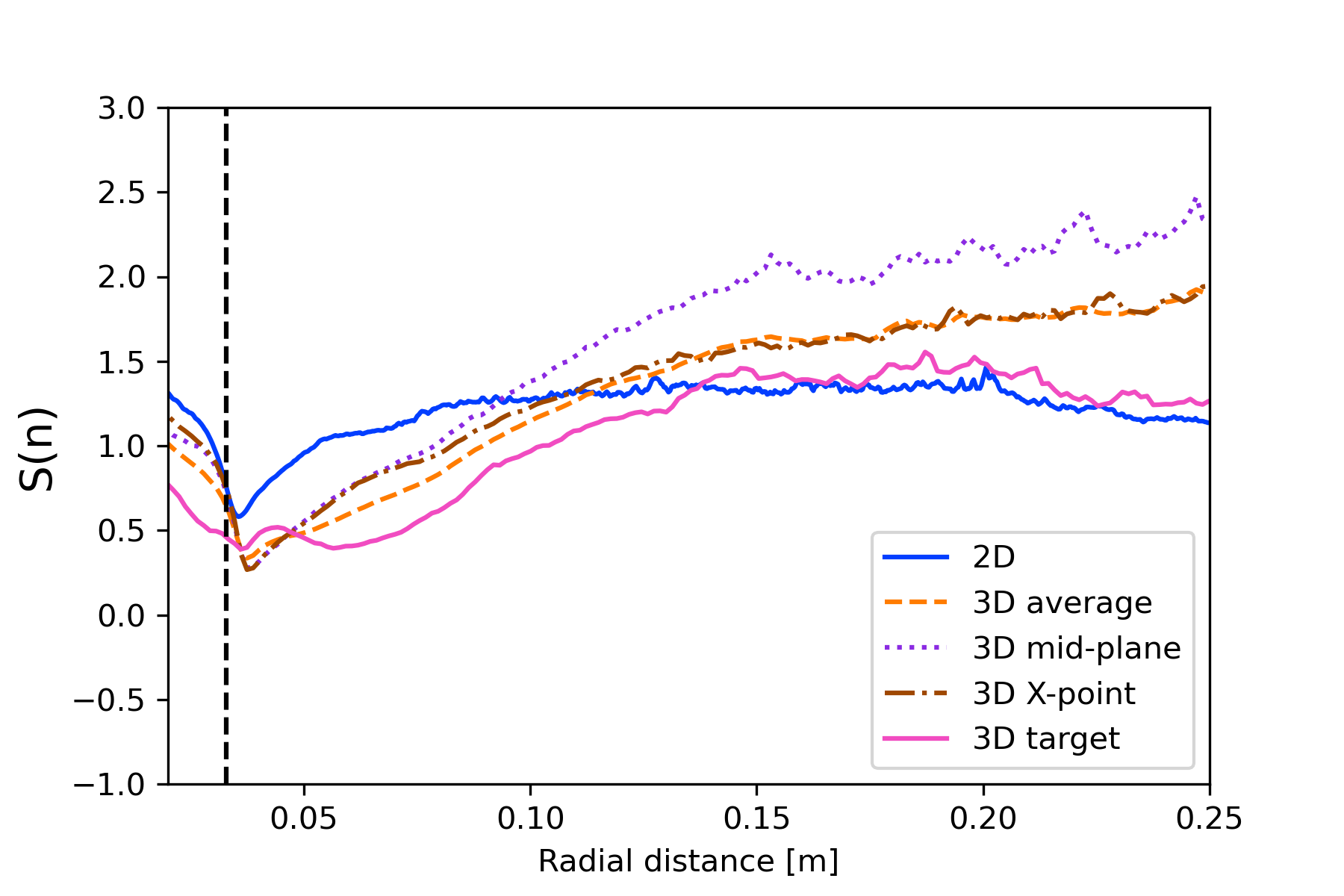}
\caption{Average radial profiles of the skewness of the normalized density fluctuations.
Averages are performed over the saturated time periods, over the binormal direction, and in the case of the line labelled ``3D average'', also over the parallel domain.
The radial location of the density and temperature sources is shown by the black dotted line.}
\label{fig:nskewprofiles}
\end{figure}

\begin{figure}[ht]
\includegraphics[width=\columnwidth,trim={0.2cm 0.25cm 1cm 1.2cm},clip]{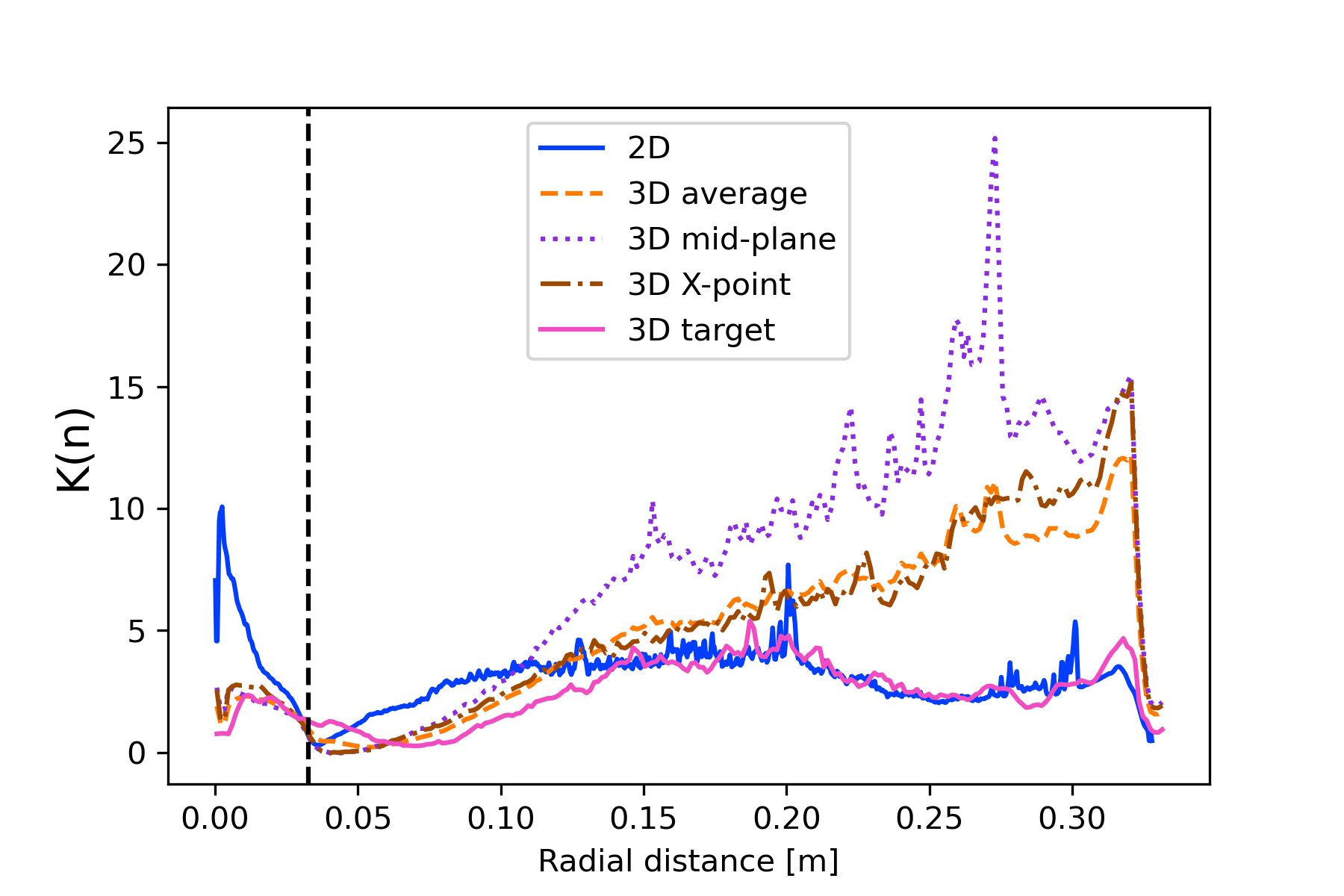}
\caption{Average radial profiles of the kurtosis of normalized density fluctuations.
Averages are performed over the saturated time periods, over the binormal direction, and in the case of the line labelled ``3D average'', also over the parallel domain.
The radial location of the density and temperature sources is shown by the black dotted line.}
\label{fig:nkurtprofiles}
\end{figure}

\begin{figure}[ht]
\includegraphics[width=\columnwidth,trim={0.2cm 0.3cm 0.0cm 0.2cm},clip]{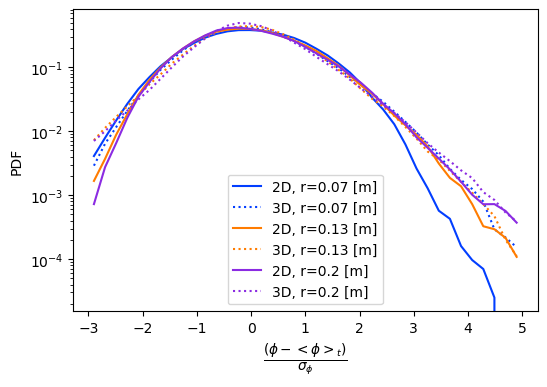}
\caption{Probability distribution functions of the fluctuations of the potential away from its mean value, plotted for different radial positions.
The distributions are averaged over the binormal direction, and the statistics for the 3D case are are taken at the mid-plane.
There is no significant systematic difference in the fluctuation distribution between the 2D and 3D simulations.
}
\label{fig:phipdfs}
\end{figure}

Figures \ref{fig:nstddevprofiles}, \ref{fig:nskewprofiles} and \ref{fig:nkurtprofiles} show that the standard deviation, skewness, and kurtosis of the parallel-averaged fluctuations in 3D are comparable everywhere to the fluctuations in 2D.
In the far SOL the skewness and kurtosis are relatively higher in 3D though - indicating that the 3D simulation allows a greater fraction of unusually high-density structures to persist out to the far SOL.
This qualitative correspondence appears to hold for the whole distribution, not just for the first few moments - the probability distribution functions of the fluctuations in the potential show little difference between 2D and 3D in the SOL (figure \ref{fig:phipdfs}).

This similarity is potentially encouraging for the purposes of comparison: it may indicate that the dominant perpendicular dynamics is largely unchanged by addition of parallel physics, and hence captured by the 2D model.
This would then mean it is not important to capture the parallel modes which are present in 3D but not in 2D - for example the possibility of drift-waves in the 3D simulations.

However it is also possible that the dynamics needed to recreate the ``universality'' of fluctuation spectra observed both here and in experiment are relatively common\cite{Antar2003}, and that just because a similar distribution is formed a similar underlying mechanism is not implied.

\subsubsection{Power spectra}

The power spectra can tell us about cross-scale energetic transfer, as well as forcing and dissipation scales (at least in cases where the scales are not very broad).

\begin{figure}[ht]
\includegraphics[width=\columnwidth,trim={0.2cm 0.0cm 0.cm 0cm},clip]{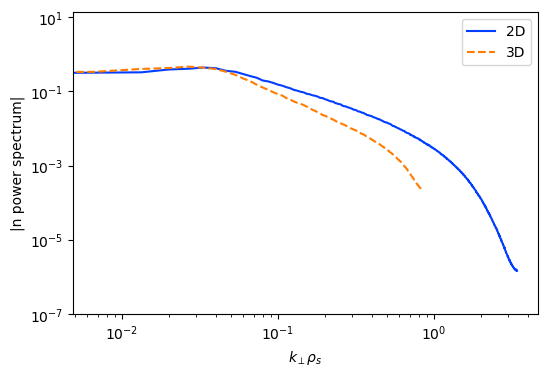}
\caption{Power spectrum of density fluctuations at different spatial scales.
In 2D the energy cascades to smaller scales before being dissipated.}
\label{fig:powerspectra}
\end{figure}

Figure \ref{fig:powerspectra} shows that dissipative mechanisms remove power from the fluctuations at a larger scale in the 3D simulation than in the 2D simulation.
This observation motivated the choice to use substantially higher resolution in the 2D simulations, to verify that the cascade region was not distorted by the dissipation region.
Nevertheless, appendix \ref{appendix:resolutionscan} shows that the choice of resolution in 3D does not significantly affect the overall results, implying that the location of the diffusion scale plays only a minor role.
The temporal Fourier spectra (not shown) exhibit a similar trend: in 2D the activity extends down to smaller temporal scales.

It is not clear what causes this effect, or whether it is physically relevant.
As Garcia et\ al. pointed out\cite{Garcia2006}, a Fourier decomposition of the perpendicular scales in the vorticity equation (eq. (30) Garcia et\ al. 2006) shows that the form of the parallel loss term affects the scales that are preferentially damped.



\subsubsection{Parallel heat fluxes}
\label{sect:heatloss}

The biggest systematic difference between the radial profiles in 2D and 3D is in the temperature profiles, so we now look at the cause of these differences in the form of the parallel heat fluxes.

\begin{figure}[ht]
\includegraphics[width=\columnwidth,trim={0.0cm 0.25cm 0.0cm 0.1},clip]{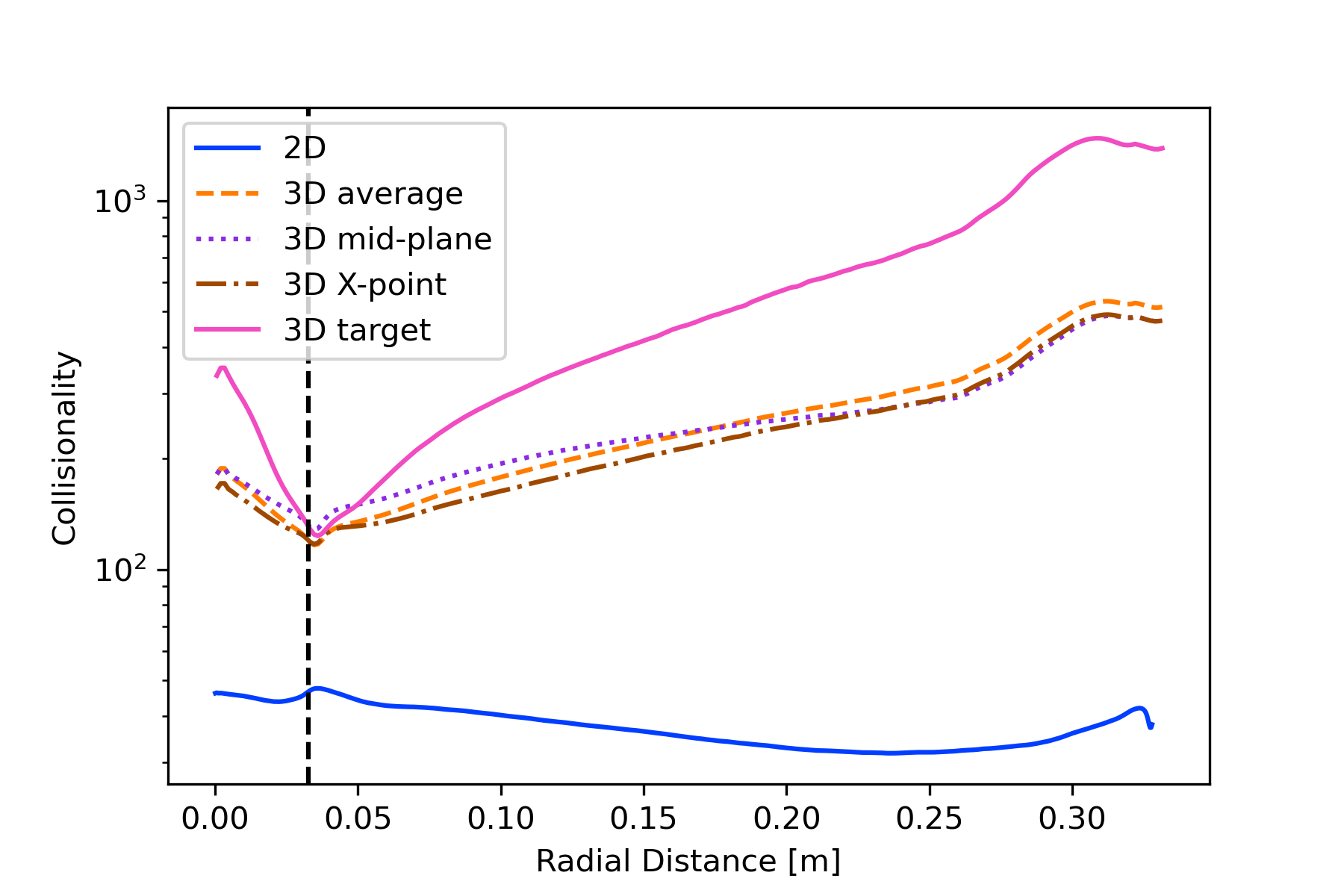}
\caption{Average radial profiles of collisionality.
Whilst the 2D simulations have $\nu^* \sim 35$ everywhere, the 3D simulations are much more collisional.
}
\label{fig:pcollisionalityprofile}
\end{figure}

\begin{figure}[ht]
\includegraphics[width=\columnwidth,trim={0.5cm 0.25cm 1cm 1},clip]{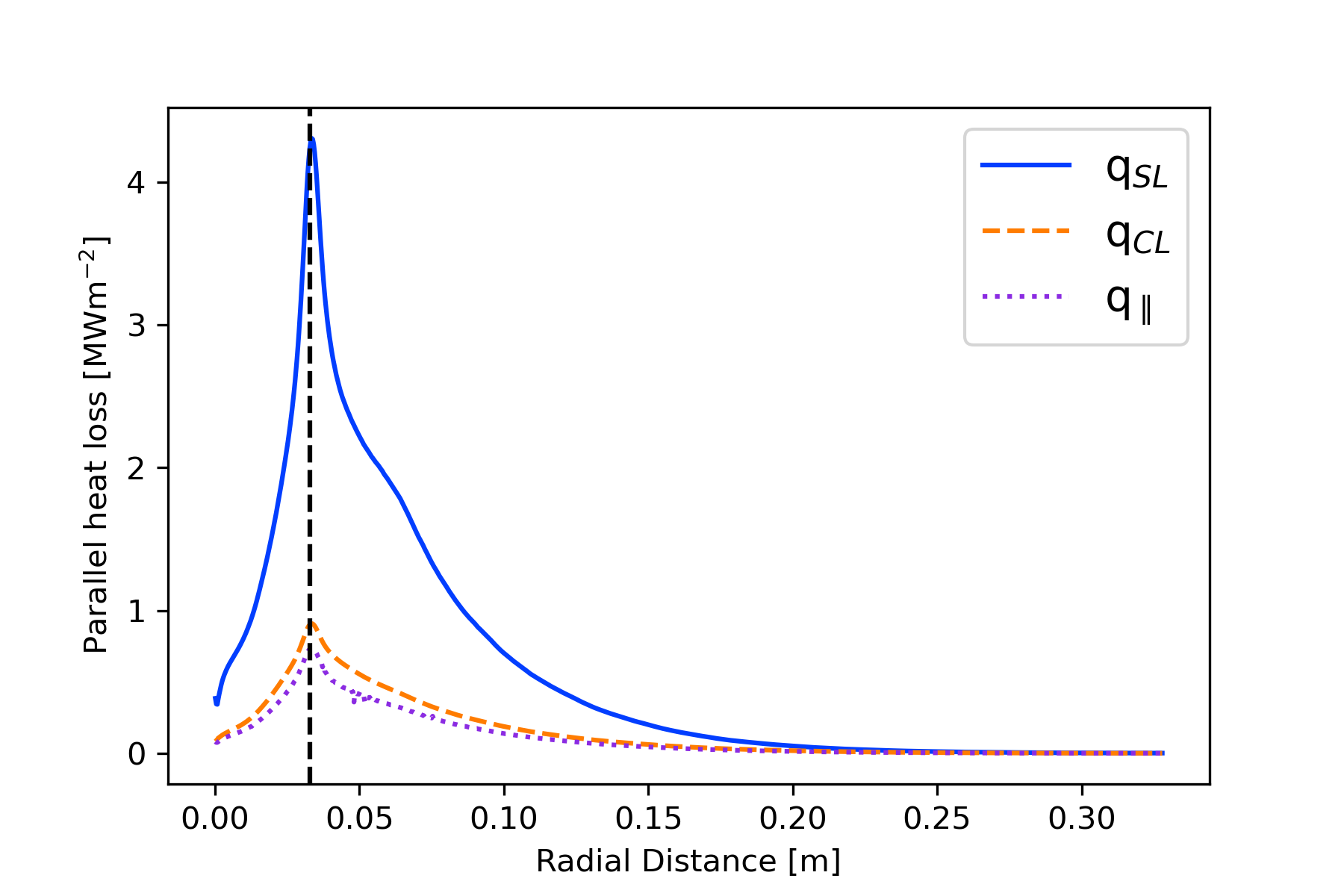}
\caption{Average radial profiles of parallel heat flux in 2D.
The harmonic averaging procedure in \eqref{eq:interpolation} means the total parallel heat flux in 2D, $q_\parallel$, is limited by the smaller value.
Here $q_\text{CL}$ is by far the smaller value, meaning this 2D simulation is in the conduction-limited regime.
The radial location of the density and temperature sources is shown by the black dotted line.}
\label{fig:heatloss2Dprofile}
\end{figure}

The collisionality can be estimated through equation (4.105) from Stangeby's textbook\cite{Stangeby2000}
\begin{equation}
    \nu^* = 1.0 \times 10^{-16}  \frac{n L_\parallel}{T^2} ,
\end{equation}
where $n$ is in m$^{-3}$, $L_\parallel$ in $m$ and $T$ in eV.
Figure \ref{fig:pcollisionalityprofile} shows that in 2D the collisionality is high everywhere, which is due to the low temperature.
The 2D parallel heat transfer is therefore very much in the conduction-limited regime (figure \ref{fig:heatloss2Dprofile}).

As an aside, because the collisionality is high, we do not need the flux-limiting term in 2D that is included by Myra\cite{Myra2011} in the harmonic average \eqref{eq:interpolation} and by Fundamenski\cite{Fundamenski2007} in an adjusted definition of parallel thermal conductivity.

\begin{figure}[ht]
\includegraphics[width=\columnwidth,trim={0.0cm 0.25cm 0.0cm 1},clip]{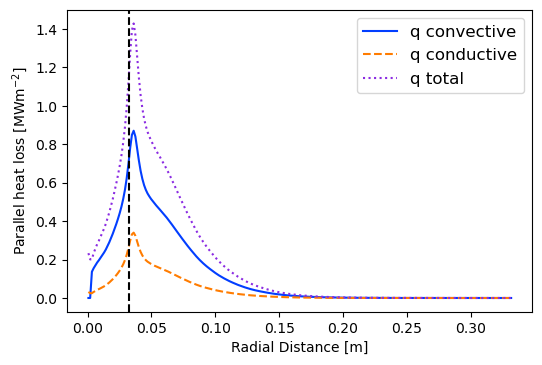}
\caption{Average radial profiles of parallel heat flux in 3D, through a single target.
The parallel transport is in a convective regime.}
\label{fig:heatloss3Dprofile}
\end{figure}

In the 3D models the average parallel heat flux is evaluated near the target in figure \ref{fig:heatloss3Dprofile}.
The peak loss at a single target is $\sim40\%$ larger than in 2D, which indicates the difference in radial temperature profiles in figure \ref{fig:Tprofiles} is due to increased parallel heat transport at the same temperature rather than due to decreased radial transport.
Decomposing the parallel transport into conductive and convective terms, again the conductive loss is smaller, which in 3D means the total loss is dominated by the convection.
Therefore whilst in 2D the conduction is the limiting quantity, in 3D we have found that the convection is more significant instead.

The condition for the dominance of conduction over convection (and hence also for the validity of the standard two-point model \cite{Stangeby2000}) can be estimated as
\begin{equation}
    \kappa_\parallel \frac{T_\text{up}}{L_\parallel} \gg n T u_\parallel.
\end{equation}
Rewriting in terms of the parallel electron mean free path $\lambda_\text{mfp}$, the Mach number $M$, the sounds speed $c_s$, and the electron thermal velocity $v_{\text{th},e}$ this becomes
\begin{equation}
    v_{\text{th},e} \frac{\lambda_\text{mfp}}{L_\parallel} \gg M c_s.
\end{equation}
Therefore in conjunction with the requirement to be in the collisional fluid limit ($\nu^* \gg 1$), the regime of validity of the two-point model (which assumes conduction dominates over convection) is a function of the local collisionality $\nu^*$ through
\begin{equation}
\label{eq:condvsconv}
    1 \ll \nu^* \ll M^{-1} \sqrt{\frac{m_i}{m_e}}.
\end{equation}
In \eqref{eq:condvsconv} the left-hand inequality is necessary to be in the fluid limit, and the right-hand inequality expresses convection being smaller than conduction.

In 3D, despite having the same total sources as in 2D, the lower temperature profile which emerges leads to a very high value of collisionality, shown in figure \ref{fig:pcollisionalityprofile}.
Figure \ref{fig:pcollisionalityprofile} shows that $\nu^* > 100$ everywhere in 3D, completely violating the right-hand inequality in \eqref{eq:condvsconv}.
This explains the presence of significant convective as well as conductive heat loss in the 3D simulations.

Conventionally the dependence on Mach number $M$ is removed from  \eqref{eq:condvsconv}, because at the target the Bohm conditions normally imply $M=1$.
However, in this case the supersonic flow near the target allows $M \sim 2$ (see section \ref{sect:leglength}), making the value of collisionality where convection becomes comparable to conduction a factor of two lower again than it would be otherwise.

Overall this means that despite choosing sources to be consistent between simulations, and values representative of the MAST SOL, the results are not in the regime of validity usually assumed: in 2D the conduction dominates but the conventional assumption of negligible convection is not well-justified, and in 3D convection dominates significantly.

\subsection{Varying divertor leg length}
\label{sect:leglength}


The parallel connection length is a key parameter in SOL physics, which sets the size of the 2D parallel loss terms, and sets the domain size in 3D.
However it is not always obvious what numerical value to use for the parallel connection length in 2D - some models use the physical length of the field-line from mid-plane to target, whilst others instead use a value intended to represent the characteristic parallel length of field-aligned structures\cite{Militello2011}.
We therefore treat this as a sensitivity parameter.

In 3D we compare 3 simulations with the same sources but different length divertor legs, the intention being to determine which lengths agree or disagree with 2D simulations with the same connection length.
The length of the leg beyond the X-point is alternatively doubled and halved relative to the baseline simulation, creating three domains with a respective $L_\parallel/L_S$ ratio of $\qty{3,2,1.5}$, where the size of the source region $L_S$ is fixed.

\begin{figure}[ht]
\includegraphics[width=\columnwidth,trim={0.2cm 0.15cm 0.2cm 0.2cm},clip]{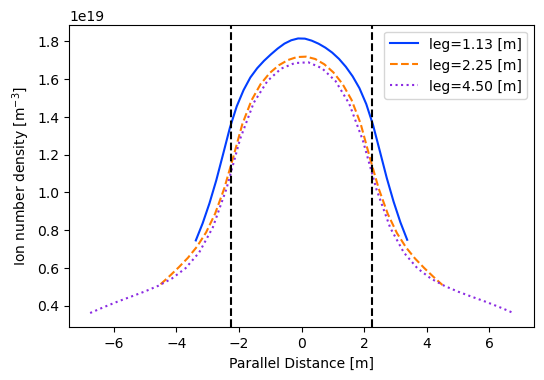}
\caption{Parallel profiles of density for 3D simulations with differing divertor leg lengths, averaged over time at radial position of 6cm (close to the particle source).
The length of the leg does not significantly affect the density at the mid-plane, or the overall profile.}
\label{fig:parallelnprofiles0.06}
\end{figure}

\begin{figure}[ht]
\includegraphics[width=\columnwidth,trim={0.2cm 0.15cm 0.2cm 0.2cm},clip]{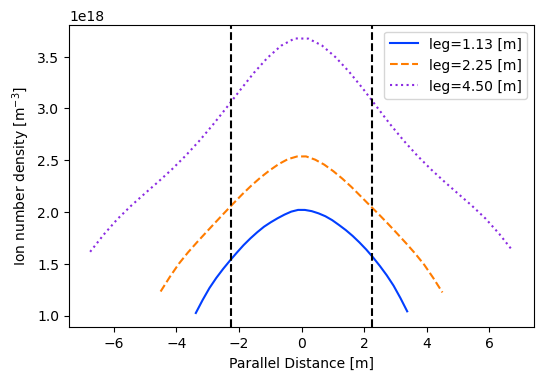}
\caption{Parallel profiles of density for 3D simulations with differing divertor leg lengths, averaged over time at radial position of 15cm (in the far SOL).
The length of the leg does not significantly affect the overall profile, but the densities are separated due to the different overall radial decay lengths.
Therefore the shorter leg is allowing higher particle flux into the target and out of the domain.}
\label{fig:parallelnprofiles0.15}
\end{figure}

\begin{figure}[ht]
\includegraphics[width=\columnwidth,trim={0.2cm 0.15cm 0.2cm 0.2cm},clip]{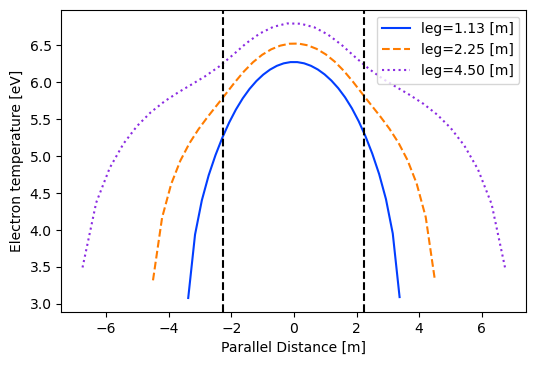}
\caption{Parallel profiles of temperature for 3D simulations with differing divertor leg lengths, averaged over time at radial position of 6cm (close to the energy source).
Whilst the temperature is peaked at mid-plane, most of the temperature loss occurs in the last 1m before the target, regardless of leg length.}
\label{fig:parallelTprofiles0.06}
\end{figure}

\begin{figure}[ht]
\includegraphics[width=\columnwidth,trim={0.2cm 0.15cm 0.2cm 0.2cm},clip]{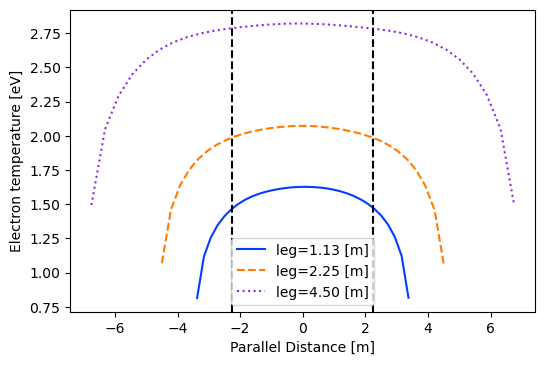}
\caption{Parallel profiles of temperature for 3D simulations with differing divertor leg lengths, averaged over time at radial position of 15cm (in the far SOL).
The temperature is mostly constant along parallel direction (figure \ref{fig:parallelTprofiles0.15}) - the drop occurs in a small region in front of the target.}
\label{fig:parallelTprofiles0.15}
\end{figure}


The average radial profiles at the mid-plane are affected: figures \ref{fig:parallelnprofiles0.06}-\ref{fig:parallelTprofiles0.15} show that the mid-plane density and temperature values are similar at the position of the sources, but separated in the far SOL.
Decay lengths of $\lambda_n=\qty{4.3,5.4,7.6}$cm and $\lambda_T=\qty{6.8,8.2,11.2}$cm are measured for $L_\parallel=\qty{1.13,2.25,4.50}$m respectively.
This dependence is weaker than linear with leg length, and a change of a factor of 2 is not sufficient to bring the temperature profile to match that observed in 2D (which had $\lambda_T=15.3$cm).
This implies the profiles have limited sensitivity to the leg length in 3D, so the agreement with the 2D model is relatively robust in that sense.

Looking specifically at the parallel variation not present in 2D, figure \ref{fig:parallelnprofiles0.06} also shows that the parallel density profile at the mid-plane is not significantly affected by changing the divertor leg length, but figure \ref{fig:parallelnprofiles0.15} shows that the densities in the far SOL are separated due to a slight difference in the different overall radial decay lengths.
Therefore the shorter leg is allowing a slightly higher particle flux into the target and out of the domain.
The continued drop of density towards the target within the leg is due to the lack of localised divertor sources or recycling in these simulations.

Figure \ref{fig:parallelTprofiles0.06} shows that while the temperature is peaked at mid-plane, most of the temperature loss occurs in the last 1m before the target, regardless of leg length.
In the far SOL the temperature is mostly constant along the parallel direction (figure \ref{fig:parallelTprofiles0.15}) - the drop still occurs in a small region in front of the target.
Like figure \ref{fig:parallelnprofiles0.15}, the profiles have separated showing a difference in overall parallel loss rate.

\begin{figure}[ht]
\includegraphics[width=\columnwidth,trim={0.2cm 0.15cm 0.2cm 0.2cm},clip]{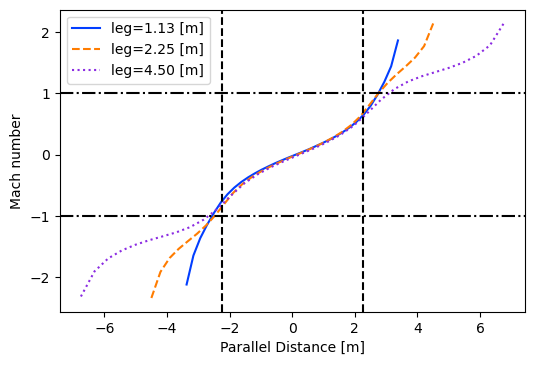}
\caption{Parallel profiles of mach number for 3D simulations with differing divertor leg lengths, averaged over time at radial position of 6cm (close to the particle source).
The length of the leg does not significantly affect the overall profile.
A transition to supersonic flow can be seen in all simulations, which occurs in the divertor leg beyond the "X-point" (marked by the vertical dotted lines).}
\label{fig:parallelmprofiles0.06}
\end{figure}

\begin{figure}[ht]
\includegraphics[width=\columnwidth,trim={0.2cm 0.15cm 0.2cm 0.2cm},clip]{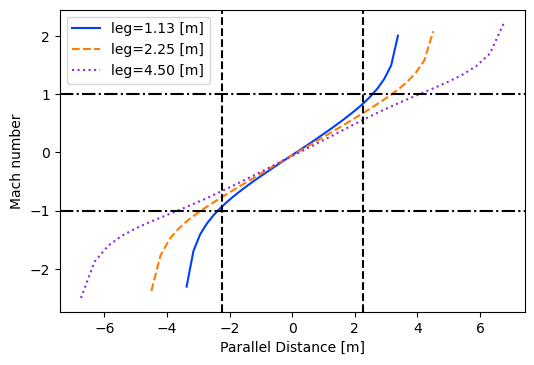}
\caption{Parallel profiles of mach number for 3D simulations with differing divertor leg lengths, averaged over time at radial position of 15cm (in the far SOL).
The length of the leg does not significantly affect the overall profile.
A transition to supersonic flow can be seen in all simulations, which occurs in the divertor leg beyond the ``X-point''.}
\label{fig:parallelmprofiles0.15}
\end{figure}

Figure \ref{fig:parallelmprofiles0.06} shows that the parallel velocity reaches sonic speeds before the target in all cases.
This result might at first appear to be inconsistent with standard nozzle theory, which derives the parallel gradient of the plasma Mach number to be (as shown in section (1.8.2.3) of \cite{Stangeby2000})
\begin{equation} \label{eq:NozzleMach}
    \dv{M}{s_\parallel} = \frac{S_n}{n c_s} \frac{(1 + M^2)}{(1 - M^2)}.
\end{equation}
However in our case $M=1$ would not correspond to a singularity in the Mach gradient because we also have $S_n=0$ at the same location ($S_n=0$ anywhere past the ``X-point'' by assumption in our geometric setup).
The mach-gradient relation \ref{eq:NozzleMach} arguably still rules out this sonic transition in the source region.
(The model used to derive \eqref{eq:NozzleMach} is isothermal, unlike our model, but figure \ref{fig:parallelTprofiles0.06} shows that the temperature only drops by $\sim 25\%$ from the mid-plane to the sonic region.)

Supersonic flow is expected in the system being studied here.
The parallel conduction of heat causes a temperature drop along the parallel direction, which lowers the local sound speed.
Meanwhile the density also drops away from the mid-plane density source.
Therefore an increasingly rarefied and cold plasma accelerates away from the mid-plane.
This model includes no momentum exchange terms to stop this flow from accelerating to supersonic speeds.
That the inclusion of momentum exchange terms is necessary to avoid cold supersonic flow is expected given that the plasma fluid has no way to exchange parallel momentum because parallel viscosity is neglected, there are no neutral collisions, and no charge exchange.
This been noted before in 1D modelling of detachment onset, such as with SD1D\cite{Dudson2019a}.

Another valid interpretation of this phenomenon is that whilst the spatially-varying parameter $S_n$ is explicitly set to zero beyond the X-point, cross-field transport between flux tubes creates an effective particle sink, which is what is actually represented in the 1D model used to derive \eqref{eq:NozzleMach}.
This sink makes the effective $S_n$ negative, so that even with $M>1$ past the X-point the plasma continues to accelerate supersonically.
As Ghendrih et\ al. \cite{Ghendrih2011} describe, ``when this particle sink prevails from the X-point region towards the target plate, one finds that transitions to supersonic flows will occur, the bifurcation point being in the vicinity of the X-point''.

This phenomenon then takes the dynamics further into a high collisionality regime, by increasing $M$ in \eqref{eq:condvsconv}, making the assumption that parallel thermal transport is only through conduction increasingly invalid.




This indicates a clear avenue for future work: a similar domain setup but with a density source in the divertor should prevent supersonic parallel flows, whilst imitating a high-recycling regime.
It would therefore be closer to an experimental reactor-relevant regime, and also include more of the key neutral effects identified as being important in 1D studies (such as \cite{Dudson2019a}).
The challenge of this model experiment would be in either (1)
deciding what shape \& size the sources should be without introducing too many arbitrary assumptions, or (2) choosing the simplest possible neutrals model that can satisfactorily model the recycling in the divertor.
A starting point for (1) might use a divertor source with a radial exponential decay length set to the typical mean free path of ions before neutralisation.


\subsection{Sensitivity to parameters in 2D}
\label{sect:sensitivity}

We also tested the sensitivity of the 2D results  with a simple parameter scan, alternately doubling or halving (a) the magnitude of all the parallel loss terms, (b) the magnitude of the particle source term, and (c) the magnitude of the energy source term.
The aim was to assess whether the effect on the profiles was linear with changing parameters, and to see if a regime of closer agreement with the 3D models existed.
For each change ((a),(b), or (c)), we looked at the effect on the time-averaged radial density and temperature profiles, for both the absolute value and the shape (by normalising to the maximum value).
Whilst we will now describe the results for all these cases, we will only show the profiles for the cases which displayed some form of non-trivial or unexpected result.



\begin{figure}[ht]
\includegraphics[width=\columnwidth,trim={0.1cm 0.20cm 0.0cm 0.0cm},clip]{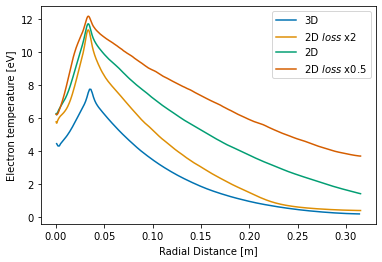}
\caption{Alternately doubling or halving the magnitude of the parallel loss terms ($n_\text{loss}$, $\Omega_\text{loss}$, $T_\text{loss}$ in \eqref{eq:STORMdensity2D} - \eqref{eq:STORMtemperature2D}) scaled the density profiles proportionally everywhere as expected (not shown), but changed the decay length of the temperature profiles without altering their maximum value.
Therefore scaling the loss terms by a overall constant factor cannot create agreement with the 3D profiles in both density and temperature.}
\label{fig:sink_size_T}
\end{figure}


Doubling or halving the magnitude of all parallel loss terms in 2D increases or decreases the average density profiles at all radial locations, as expected.
It also changes the temperature profiles, but in a way which leaves the maximum value fixed, whilst altering the decay lengths (figure \ref{fig:sink_size_T}).
Therefore the magnitude of the loss terms cannot be used on its own to tune the 2D simulations to match the 3D results in both density and temperature because increasing the losses (equivalent to a shorter $L_\parallel$) will depress the density profiles as well as the temperature decay lengths, and the temperature profiles do not change their absolute value so will not match either.

\begin{figure}[ht]
\includegraphics[width=\columnwidth,trim={0.1cm 0.20cm 0.0cm 0.0cm},clip]{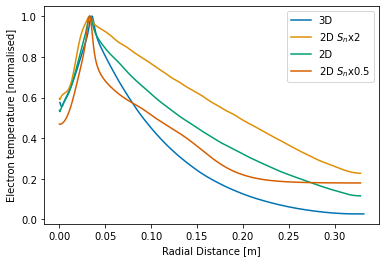}
\caption{Altering the density source $S_n$ by a factor of 2 scaled the density profiles linearly at all locations (not shown), but also altered the shape of the normalised temperature profiles relative to their maximum value.}
\label{fig:n_source_size_T_profile_normed}
\end{figure}

Altering the magnitude of the density source terms has a larger effect on the density profiles than changing the size of the loss terms does, with the absolute height of the profiles varying almost linearly with the density sources.
When normalised, the profiles are coincident, so there is no change the the decay length, as expected.
The temperature profile scales inversely with the size of the density source, which is expected because the same energy has been distributed amongst twice as many particles in the same time period.
However, increasing the density sources also broadens the normalised temperature profile, shown in figure \ref{fig:n_source_size_T_profile_normed}.

\begin{figure}[ht]
\includegraphics[width=\columnwidth,trim={0.1cm 0.20cm 0.0cm 0.0cm},clip]{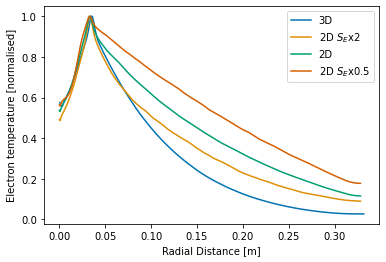}
\caption{Scaling the energy source $S_E$ by a factor of 2 scales the absolute value of the temperature profiles linearly as expected (not shown), but also has the effect of broadening the temperature profiles.}
\label{fig:E_source_size_T_profile_normed}
\end{figure}

\begin{figure}[ht]
\includegraphics[width=\columnwidth,trim={0.1cm 0.20cm 0.0cm 0.0cm},clip]{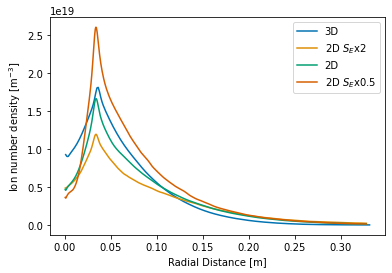}
\caption{Scaling the energy source $S_E$ by a factor of 2 scales the absolute value of the density profiles inversely.}
\label{fig:E_source_size_n_profile}
\end{figure}

Scaling the energy source terms causes the absolute value of the temperature profile to vary linearly as expected, but changes the normalised shape of the profiles (figure \ref{fig:E_source_size_T_profile_normed}), with smaller energy input creating broader profiles.
Therefore, as the 2D temperature profile is broader than the 3D one, and scaling the energy sources does not change the shape, the discrepancy between the 2D and 3D temperature profiles cannot be resolved by simply scaling the energy sources in 2D.
The density profile scales inversely with energy source (figure \ref{fig:E_source_size_n_profile}), but with no change to the normalised shape.





Since decreasing the energy source brings the absolute temperature closer to the 3D profile but makes the 2D density increase, we also tried decreasing both the density and the energy source terms (again by a factor of 2).
We found that whilst the temperature profiles now matched 3D, the density profile was everywhere lower.

We conclude that over a range of a factor of 4 changing the magnitude of the loss and source terms in the 2D simulation is not sufficient to easily recreate the average 3D profiles of both density and temperature simultaneously.

\section{Conclusions}



There is extensive literature on two-dimensional simulations of SOL drift-plane turbulence, including experimental validation.
We have taken a first-principles approach to evaluating the approximations used to truncate the models to 2D, comparing the 2D case with 3D simulations set up so as to form the closest possible analogues of the 2D models.


The 2D model successfully replicates the mean density profile of the 3D models with matched sources and no parameter tuning.
The 2D model also reproduces the fluctuation statistics of the 3D model well, and the results are robust to changing the length of the divertor leg in 3D or the source and sink sizes in 2D.

One systematic difference was a broadening of the mean temperature profile in 2D compared to in 3D.
The reason for this difference is an interplay of issues caused by two reasonable-seeming assumptions in the model setup. 
In particular, in very high collisionality conditions, the assumptions used to close 2D models are not valid, because high enough collisionality suppresses thermal conduction so that convection becomes important.
Further, the source setup used in the parallel direction for the 3D simulations causes supersonic flows beyond the X-point, which make the thermal convection compete with conduction at lower collisionality than would otherwise be the case.
Since this highly supersonic flow is not typically observed in experiments, we suggest that future work always include particle sources in the divertor leg.

The good qualitative agreement between these two types of reduced models provides a basis for interpreting 2D and 3D models relative to one another,  our results also motivate future work.
Our results open up avenues for future work moving the 3D models step by step towards more detailed representations of tokamak divertors, in particular using a larger power source (to obtain a higher temperature and hence lower collisionality) and a divertor density source (to keep the parallel flows subsonic), would provide a more consistently analogous system.



\subsection*{Acknowledgements}

The corresponding author was financially supported by the EPSRC Centre for Doctoral Training in the Science and Technology of Fusion Energy [Grant Number EP/L01663X/1], as well as an iCASE award from CCFE.
This work has also been part-funded by the RCUK Energy Programme [grant number EP/T012250/1].
All simulations presented were performed on the CINECA Marconi supercomputer within the framework of the FUA33\_SOLBOUT3 and FUA34\_SOLBOUT4 projects.

\appendix

\section{Boundary Conditions}
\label{appendix:bcs}

At the target plates the first-principles Bohm boundary conditions set $U$, $V$ and $q_\parallel$ through \eqref{eq:sheathbcs}-\eqref{eq:sheathbcsq}.
For the other variables (i.e. $\phi$, $n$, $T$, $\Omega$), free boundary conditions are used at the targets.

Unfortunately there is no first-principles theory describing how a fluid plasma model interacts with the vessel wall at the radial boundaries available in the literature, and we also wish to represent the core-SOL interaction with as few assumptions as possible.
For the fluid variables $n$, $T$, $\Omega$, $U$ and $V$ Neumann boundary conditions are used in order to allow the variables to ``float'' and hence be determined by the balance of radial turbulent transport and parallel loss.

However, the electric potential must still be constrained in order to satisfy Laplace's equation \eqref{eq:STORMpotential}, so we employ the same ad hoc approach used in \cite{Riva2019}.
We employ ``evolving boundary conditions'' in which the potential everywhere on the boundary is set to the mean value of the potential on that boundary averaged over the preceding time period $\tau$.
Formally this means we set $\phi$ on the inner and outer radial boundaries $x_i$ and $x_o$ through
\begin{equation}
\begin{split}
    \phi(x = x_i) &= \langle \phi(x = x_i + \Delta x_i / 2) \rangle_{z, t \in [(j-1)\tau, j\tau]} \\
    \phi(x = x_0) &= \langle \phi(x = x_0 - \Delta x_o / 2) \rangle_{z, t \in [(j-1)\tau, j\tau]}
\end{split}
\end{equation}
for all $t \in [j\tau, (j+1)\tau]$, where $j \in \mathbb{Z}^+$, and $\langle - \rangle_{z, t \in [(j-1)\tau, j\tau]}$ denotes a binormal- and time-average over the time interval $[(j-1)\tau,  j\tau]$.
$\tau$ is an input parameter which sets the length of preceding time over which the $\phi$ values are averaged, and all the simulations presented here used $\tau = 50/\Omega_{i,0}.$
In 2D the same conditions are used at the radial boundaries, and the targets are treated through the closure approximation described in section \ref{sect:closure2D}.
This approach allows us to constrain fluctuations in the potential at the boundaries without fixing it at any specific arbitrary value.
In practice the potential at the boundaries would display an initial transient phase, before converging on a specific value that depended on the particular simulation.
All analysis was only performed on data obtained after the potential had converged.
Nevertheless, the impact of this choice of boundary conditions upon the simulations was mitigated further by choosing a domain with large radial extent, and by excluding the regions near the boundary during parts of the analysis.

\section{Scan in Grid Resolution}
\label{appendix:resolutionscan}

To test the impact of the choice of grid resolution on the 3D simulations used for results presented in sections \ref{sect:Comparison2Dvs3D} and \ref{sect:leglength}, additional 3D simulations with different perpendicular grid resolutions were run.

The baseline simulation used for the main results had $240 \times 32 \times 256$ grid points along $\qty(x,y,z)$, which was supplemented by a lower resolution simulation with $120 \times 32 \times 128$ grid points, and a higher resolution one with $480 \times 32 \times 512$ grid points.

\begin{figure}[ht]
\includegraphics[width=\columnwidth,trim={0.1cm 0.2cm 0.5cm 0.1cm},clip]{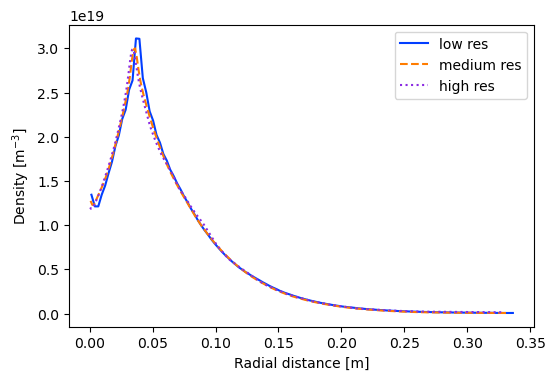}
\caption{Averaged profiles of density for 3D simulations with three different perpendicular resolutions, showing negligible changes.}
\label{fig:profilesresolutionscan}
\end{figure}

Figure \ref{fig:profilesresolutionscan} shows that the change in resolution has negligible effect on the average density profile (a similar lack of variation is seen in all other averaged variables).

There is some difference in the power spectra of the density fluctuations: the lower resolution simulations show an small ``arching'' of the power spectrum at smaller spatial scales, but the difference between the medium resolution simulation used and the higher resolution one is small.

\begin{figure}[ht]
\includegraphics[width=\columnwidth,trim={0.1cm 0.2cm 0.2cm 0.1cm},clip]{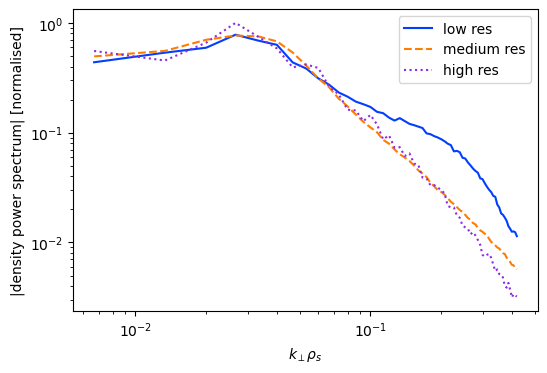}
\caption{Power spectra of density fluctuations compared across three 3D simulations with different perpendicular resolutions.
There is some difference for the lower resolutions, but this reduces to only a small difference when comparing the medium resolution case (the one used for the main results) to the high resolution one.
}
\label{fig:spectraresolutionscan}
\end{figure}

\bibliography{thesis} 

\begin{thebibliography}{39}%
\makeatletter
\providecommand \@ifxundefined [1]{%
 \@ifx{#1\undefined}
}%
\providecommand \@ifnum [1]{%
 \ifnum #1\expandafter \@firstoftwo
 \else \expandafter \@secondoftwo
 \fi
}%
\providecommand \@ifx [1]{%
 \ifx #1\expandafter \@firstoftwo
 \else \expandafter \@secondoftwo
 \fi
}%
\providecommand \natexlab [1]{#1}%
\providecommand \enquote  [1]{``#1''}%
\providecommand \bibnamefont  [1]{#1}%
\providecommand \bibfnamefont [1]{#1}%
\providecommand \citenamefont [1]{#1}%
\providecommand \href@noop [0]{\@secondoftwo}%
\providecommand \href [0]{\begingroup \@sanitize@url \@href}%
\providecommand \@href[1]{\@@startlink{#1}\@@href}%
\providecommand \@@href[1]{\endgroup#1\@@endlink}%
\providecommand \@sanitize@url [0]{\catcode `\\12\catcode `\$12\catcode
  `\&12\catcode `\#12\catcode `\^12\catcode `\_12\catcode `\%12\relax}%
\providecommand \@@startlink[1]{}%
\providecommand \@@endlink[0]{}%
\providecommand \url  [0]{\begingroup\@sanitize@url \@url }%
\providecommand \@url [1]{\endgroup\@href {#1}{\urlprefix }}%
\providecommand \urlprefix  [0]{URL }%
\providecommand \Eprint [0]{\href }%
\providecommand \doibase [0]{http://dx.doi.org/}%
\providecommand \selectlanguage [0]{\@gobble}%
\providecommand \bibinfo  [0]{\@secondoftwo}%
\providecommand \bibfield  [0]{\@secondoftwo}%
\providecommand \translation [1]{[#1]}%
\providecommand \BibitemOpen [0]{}%
\providecommand \bibitemStop [0]{}%
\providecommand \bibitemNoStop [0]{.\EOS\space}%
\providecommand \EOS [0]{\spacefactor3000\relax}%
\providecommand \BibitemShut  [1]{\csname bibitem#1\endcsname}%
\let\auto@bib@innerbib\@empty
\bibitem [{\citenamefont {Krasheninnikov}(2001)}]{Krasheninnikov2001}%
  \BibitemOpen
  \bibfield  {author} {\bibinfo {author} {\bibfnamefont {S.~I.}\ \bibnamefont
  {Krasheninnikov}},\ }\href {\doibase 10.1016/S0375-9601(01)00252-3}
  {\bibfield  {journal} {\bibinfo  {journal} {Physics Letters, Section A:
  General, Atomic and Solid State Physics}\ }\textbf {\bibinfo {volume}
  {283}},\ \bibinfo {pages} {368} (\bibinfo {year} {2001})}\BibitemShut
  {NoStop}%
\bibitem [{\citenamefont {Garcia}\ \emph {et~al.}(2005)\citenamefont {Garcia},
  \citenamefont {Naulin}, \citenamefont {Nielsen},\ and\ \citenamefont
  {Rasmussen}}]{Garcia2005}%
  \BibitemOpen
  \bibfield  {author} {\bibinfo {author} {\bibfnamefont {O.~E.}\ \bibnamefont
  {Garcia}}, \bibinfo {author} {\bibfnamefont {V.}~\bibnamefont {Naulin}},
  \bibinfo {author} {\bibfnamefont {A.~H.}\ \bibnamefont {Nielsen}}, \ and\
  \bibinfo {author} {\bibfnamefont {J.~J.}\ \bibnamefont {Rasmussen}},\ }\href
  {\doibase 10.1063/1.1925617} {\bibfield  {journal} {\bibinfo  {journal}
  {Physics of Plasmas}\ }\textbf {\bibinfo {volume} {12}},\ \bibinfo {pages}
  {1} (\bibinfo {year} {2005})}\BibitemShut {NoStop}%
\bibitem [{\citenamefont {Militello}\ \emph {et~al.}(2012)\citenamefont
  {Militello}, \citenamefont {Naulin}, \citenamefont {Nielsen}, \citenamefont
  {Tamain}, \citenamefont {Fundamenski}, \citenamefont {Garcia}, \citenamefont
  {Yan}, \citenamefont {Xu}, \citenamefont {Ricci}, \citenamefont {Halpern},
  \citenamefont {Jolliet}, \citenamefont {Pitts}, \citenamefont {Matthews},
  \citenamefont {Ahn}, \citenamefont {Counsell}, \citenamefont {Kirk},\ and\
  \citenamefont {Havl{\'{i}}kov{\'{a}}}}]{Militello2012}%
  \BibitemOpen
  \bibfield  {author} {\bibinfo {author} {\bibfnamefont {F.}~\bibnamefont
  {Militello}}, \bibinfo {author} {\bibfnamefont {V.}~\bibnamefont {Naulin}},
  \bibinfo {author} {\bibfnamefont {A.~H.}\ \bibnamefont {Nielsen}}, \bibinfo
  {author} {\bibfnamefont {P.}~\bibnamefont {Tamain}}, \bibinfo {author}
  {\bibfnamefont {W.}~\bibnamefont {Fundamenski}}, \bibinfo {author}
  {\bibfnamefont {O.~E.}\ \bibnamefont {Garcia}}, \bibinfo {author}
  {\bibfnamefont {N.}~\bibnamefont {Yan}}, \bibinfo {author} {\bibfnamefont
  {G.~S.}\ \bibnamefont {Xu}}, \bibinfo {author} {\bibfnamefont
  {P.}~\bibnamefont {Ricci}}, \bibinfo {author} {\bibfnamefont {F.~D.}\
  \bibnamefont {Halpern}}, \bibinfo {author} {\bibfnamefont {S.}~\bibnamefont
  {Jolliet}}, \bibinfo {author} {\bibfnamefont {R.~A.}\ \bibnamefont {Pitts}},
  \bibinfo {author} {\bibfnamefont {G.~F.}\ \bibnamefont {Matthews}}, \bibinfo
  {author} {\bibfnamefont {J.-W.}\ \bibnamefont {Ahn}}, \bibinfo {author}
  {\bibfnamefont {G.~F.}\ \bibnamefont {Counsell}}, \bibinfo {author}
  {\bibfnamefont {A.}~\bibnamefont {Kirk}}, \ and\ \bibinfo {author}
  {\bibfnamefont {E.}~\bibnamefont {Havl{\'{i}}kov{\'{a}}}},\ }\href {\doibase
  10.1088/0741-3335/54/9/095011} {\bibfield  {journal} {\bibinfo  {journal}
  {Plasma Phys. Control. Fusion}\ }\textbf {\bibinfo {volume} {54}},\ \bibinfo
  {pages} {95011} (\bibinfo {year} {2012})}\BibitemShut {NoStop}%
\bibitem [{\citenamefont {Russell}\ \emph {et~al.}(2009)\citenamefont
  {Russell}, \citenamefont {Myra},\ and\ \citenamefont
  {D'Ippolito}}]{Russell2009}%
  \BibitemOpen
  \bibfield  {author} {\bibinfo {author} {\bibfnamefont {D.~A.}\ \bibnamefont
  {Russell}}, \bibinfo {author} {\bibfnamefont {J.~R.}\ \bibnamefont {Myra}}, \
  and\ \bibinfo {author} {\bibfnamefont {D.~A.}\ \bibnamefont {D'Ippolito}},\
  }\href {\doibase 10.1063/1.3270051} {\bibfield  {journal} {\bibinfo
  {journal} {Physics of Plasmas}\ }\textbf {\bibinfo {volume} {16}} (\bibinfo
  {year} {2009}),\ 10.1063/1.3270051}\BibitemShut {NoStop}%
\bibitem [{\citenamefont {Myra}\ \emph
  {et~al.}(2011{\natexlab{a}})\citenamefont {Myra}, \citenamefont {Russell},
  \citenamefont {D'Ippolito}, \citenamefont {Ahn}, \citenamefont {Maingi},
  \citenamefont {Maqueda}, \citenamefont {Lundberg}, \citenamefont {Stotler},
  \citenamefont {Zweben}, \citenamefont {Boedo},\ and\ \citenamefont
  {Umansky}}]{Myra2011}%
  \BibitemOpen
  \bibfield  {author} {\bibinfo {author} {\bibfnamefont {J.~R.}\ \bibnamefont
  {Myra}}, \bibinfo {author} {\bibfnamefont {D.~A.}\ \bibnamefont {Russell}},
  \bibinfo {author} {\bibfnamefont {D.~A.}\ \bibnamefont {D'Ippolito}},
  \bibinfo {author} {\bibfnamefont {J.~W.}\ \bibnamefont {Ahn}}, \bibinfo
  {author} {\bibfnamefont {R.}~\bibnamefont {Maingi}}, \bibinfo {author}
  {\bibfnamefont {R.~J.}\ \bibnamefont {Maqueda}}, \bibinfo {author}
  {\bibfnamefont {D.~P.}\ \bibnamefont {Lundberg}}, \bibinfo {author}
  {\bibfnamefont {D.~P.}\ \bibnamefont {Stotler}}, \bibinfo {author}
  {\bibfnamefont {S.~J.}\ \bibnamefont {Zweben}}, \bibinfo {author}
  {\bibfnamefont {J.}~\bibnamefont {Boedo}}, \ and\ \bibinfo {author}
  {\bibfnamefont {M.}~\bibnamefont {Umansky}},\ }\href {\doibase
  10.1063/1.3526676} {\bibfield  {journal} {\bibinfo  {journal} {Physics of
  Plasmas}\ }\textbf {\bibinfo {volume} {18}} (\bibinfo {year}
  {2011}{\natexlab{a}}),\ 10.1063/1.3526676}\BibitemShut {NoStop}%
\bibitem [{\citenamefont {Easy}\ \emph {et~al.}(2014)\citenamefont {Easy},
  \citenamefont {Militello}, \citenamefont {Omotani}, \citenamefont {Dudson},
  \citenamefont {Havlikova}, \citenamefont {Tamain}, \citenamefont {Naulin},\
  and\ \citenamefont {Nielsen}}]{Easy2014}%
  \BibitemOpen
  \bibfield  {author} {\bibinfo {author} {\bibfnamefont {L.}~\bibnamefont
  {Easy}}, \bibinfo {author} {\bibfnamefont {F.}~\bibnamefont {Militello}},
  \bibinfo {author} {\bibfnamefont {J.}~\bibnamefont {Omotani}}, \bibinfo
  {author} {\bibfnamefont {B.}~\bibnamefont {Dudson}}, \bibinfo {author}
  {\bibfnamefont {E.}~\bibnamefont {Havlikova}}, \bibinfo {author}
  {\bibfnamefont {P.}~\bibnamefont {Tamain}}, \bibinfo {author} {\bibfnamefont
  {V.}~\bibnamefont {Naulin}}, \ and\ \bibinfo {author} {\bibfnamefont {A.~H.}\
  \bibnamefont {Nielsen}},\ }\href {\doibase 10.1063/1.4904207} {\bibfield
  {journal} {\bibinfo  {journal} {Physics of Plasmas}\ }\textbf {\bibinfo
  {volume} {21}} (\bibinfo {year} {2014}),\ 10.1063/1.4904207},\ \Eprint
  {http://arxiv.org/abs/1410.2137} {arXiv:1410.2137} \BibitemShut {NoStop}%
\bibitem [{\citenamefont {Easy}(2016)}]{Easy2016}%
  \BibitemOpen
  \bibfield  {author} {\bibinfo {author} {\bibfnamefont {L.}~\bibnamefont
  {Easy}},\ }\emph {\bibinfo {title} {{Three Dimensional Simulations of
  Scrape-Off Layer Filaments}}},\ \href@noop {} {Ph.D. thesis},\ \bibinfo
  {school} {University of York} (\bibinfo {year} {2016})\BibitemShut {NoStop}%
\bibitem [{\citenamefont {Bisai}\ \emph {et~al.}(2005)\citenamefont {Bisai},
  \citenamefont {Das}, \citenamefont {Deshpande}, \citenamefont {Jha},
  \citenamefont {Kaw}, \citenamefont {Sen},\ and\ \citenamefont
  {Singh}}]{Bisai2005}%
  \BibitemOpen
  \bibfield  {author} {\bibinfo {author} {\bibfnamefont {N.}~\bibnamefont
  {Bisai}}, \bibinfo {author} {\bibfnamefont {A.}~\bibnamefont {Das}}, \bibinfo
  {author} {\bibfnamefont {S.}~\bibnamefont {Deshpande}}, \bibinfo {author}
  {\bibfnamefont {R.}~\bibnamefont {Jha}}, \bibinfo {author} {\bibfnamefont
  {P.}~\bibnamefont {Kaw}}, \bibinfo {author} {\bibfnamefont {A.}~\bibnamefont
  {Sen}}, \ and\ \bibinfo {author} {\bibfnamefont {R.}~\bibnamefont {Singh}},\
  }\href {\doibase 10.1063/1.2083791} {\bibfield  {journal} {\bibinfo
  {journal} {Physics of Plasmas}\ }\textbf {\bibinfo {volume} {12}},\ \bibinfo
  {pages} {1} (\bibinfo {year} {2005})}\BibitemShut {NoStop}%
\bibitem [{\citenamefont {Halpern}\ \emph {et~al.}(2016)\citenamefont
  {Halpern}, \citenamefont {Ricci}, \citenamefont {Jolliet}, \citenamefont
  {Loizu}, \citenamefont {Morales}, \citenamefont {Mosetto}, \citenamefont
  {Musil}, \citenamefont {Riva}, \citenamefont {Tran},\ and\ \citenamefont
  {Wersal}}]{Halpern2016}%
  \BibitemOpen
  \bibfield  {author} {\bibinfo {author} {\bibfnamefont {F.~D.}\ \bibnamefont
  {Halpern}}, \bibinfo {author} {\bibfnamefont {P.}~\bibnamefont {Ricci}},
  \bibinfo {author} {\bibfnamefont {S.}~\bibnamefont {Jolliet}}, \bibinfo
  {author} {\bibfnamefont {J.}~\bibnamefont {Loizu}}, \bibinfo {author}
  {\bibfnamefont {J.}~\bibnamefont {Morales}}, \bibinfo {author} {\bibfnamefont
  {A.}~\bibnamefont {Mosetto}}, \bibinfo {author} {\bibfnamefont
  {F.}~\bibnamefont {Musil}}, \bibinfo {author} {\bibfnamefont
  {F.}~\bibnamefont {Riva}}, \bibinfo {author} {\bibfnamefont {T.~M.}\
  \bibnamefont {Tran}}, \ and\ \bibinfo {author} {\bibfnamefont
  {C.}~\bibnamefont {Wersal}},\ }\href {\doibase 10.1016/j.jcp.2016.03.040}
  {\bibfield  {journal} {\bibinfo  {journal} {Journal of Computational
  Physics}\ }\textbf {\bibinfo {volume} {315}},\ \bibinfo {pages} {388}
  (\bibinfo {year} {2016})}\BibitemShut {NoStop}%
\bibitem [{\citenamefont {Umansky}\ \emph {et~al.}(2019)\citenamefont
  {Umansky}, \citenamefont {Cohen},\ and\ \citenamefont {Myra}}]{Umansky2019}%
  \BibitemOpen
  \bibfield  {author} {\bibinfo {author} {\bibfnamefont {M.}~\bibnamefont
  {Umansky}}, \bibinfo {author} {\bibfnamefont {B.}~\bibnamefont {Cohen}}, \
  and\ \bibinfo {author} {\bibfnamefont {J.}~\bibnamefont {Myra}},\ }in\
  \href@noop {} {\emph {\bibinfo {booktitle} {DPP19 Meeting of The American
  Physical Society}}}\ (\bibinfo {year} {2019})\BibitemShut {NoStop}%
\bibitem [{\citenamefont {Tamain}\ \emph {et~al.}(2016)\citenamefont {Tamain},
  \citenamefont {Bufferand}, \citenamefont {Ciraolo}, \citenamefont {Colin},
  \citenamefont {Galassi}, \citenamefont {Ghendrih}, \citenamefont
  {Schwander},\ and\ \citenamefont {Serre}}]{Tamain2016}%
  \BibitemOpen
  \bibfield  {author} {\bibinfo {author} {\bibfnamefont {P.}~\bibnamefont
  {Tamain}}, \bibinfo {author} {\bibfnamefont {H.}~\bibnamefont {Bufferand}},
  \bibinfo {author} {\bibfnamefont {G.}~\bibnamefont {Ciraolo}}, \bibinfo
  {author} {\bibfnamefont {C.}~\bibnamefont {Colin}}, \bibinfo {author}
  {\bibfnamefont {D.}~\bibnamefont {Galassi}}, \bibinfo {author} {\bibfnamefont
  {P.}~\bibnamefont {Ghendrih}}, \bibinfo {author} {\bibfnamefont
  {F.}~\bibnamefont {Schwander}}, \ and\ \bibinfo {author} {\bibfnamefont
  {E.}~\bibnamefont {Serre}},\ }\href {\doibase 10.1016/j.jcp.2016.05.038}
  {\bibfield  {journal} {\bibinfo  {journal} {Journal of Computational
  Physics}\ }\textbf {\bibinfo {volume} {321}},\ \bibinfo {pages} {606}
  (\bibinfo {year} {2016})}\BibitemShut {NoStop}%
\bibitem [{\citenamefont {Stegmeir}\ \emph {et~al.}(2018)\citenamefont
  {Stegmeir}, \citenamefont {Coster}, \citenamefont {Ross}, \citenamefont
  {Maj}, \citenamefont {Lackner},\ and\ \citenamefont {Poli}}]{Stegmeir2018}%
  \BibitemOpen
  \bibfield  {author} {\bibinfo {author} {\bibfnamefont {A.}~\bibnamefont
  {Stegmeir}}, \bibinfo {author} {\bibfnamefont {D.}~\bibnamefont {Coster}},
  \bibinfo {author} {\bibfnamefont {A.}~\bibnamefont {Ross}}, \bibinfo {author}
  {\bibfnamefont {O.}~\bibnamefont {Maj}}, \bibinfo {author} {\bibfnamefont
  {K.}~\bibnamefont {Lackner}}, \ and\ \bibinfo {author} {\bibfnamefont
  {E.}~\bibnamefont {Poli}},\ }\href@noop {} {\bibfield  {journal} {\bibinfo
  {journal} {Plasma Physics and Controlled Fusion}\ }\textbf {\bibinfo {volume}
  {60}} (\bibinfo {year} {2018})}\BibitemShut {NoStop}%
\bibitem [{\citenamefont {Dudson}\ and\ \citenamefont
  {Leddy}(2017)}]{Dudson2017}%
  \BibitemOpen
  \bibfield  {author} {\bibinfo {author} {\bibfnamefont {B.~D.}\ \bibnamefont
  {Dudson}}\ and\ \bibinfo {author} {\bibfnamefont {J.}~\bibnamefont {Leddy}},\
  }\href@noop {} {\bibfield  {journal} {\bibinfo  {journal} {Plasma Physics and
  Controlled Fusion}\ }\textbf {\bibinfo {volume} {59}} (\bibinfo {year}
  {2017})}\BibitemShut {NoStop}%
\bibitem [{\citenamefont {Sykes}\ \emph {et~al.}(2001)\citenamefont {Sykes},
  \citenamefont {Ahn}, \citenamefont {Akers}, \citenamefont {Arends},
  \citenamefont {Carolan}, \citenamefont {Counsell}, \citenamefont {Fielding},
  \citenamefont {Gryaznevich}, \citenamefont {Martin}, \citenamefont {Price},
  \citenamefont {Roach}, \citenamefont {Shevchenko}, \citenamefont
  {Tournianski}, \citenamefont {Valovic}, \citenamefont {Walsh},\ and\
  \citenamefont {Wilson}}]{Sykes2001}%
  \BibitemOpen
  \bibfield  {author} {\bibinfo {author} {\bibfnamefont {A.}~\bibnamefont
  {Sykes}}, \bibinfo {author} {\bibfnamefont {J.~W.}\ \bibnamefont {Ahn}},
  \bibinfo {author} {\bibfnamefont {R.}~\bibnamefont {Akers}}, \bibinfo
  {author} {\bibfnamefont {E.}~\bibnamefont {Arends}}, \bibinfo {author}
  {\bibfnamefont {P.~G.}\ \bibnamefont {Carolan}}, \bibinfo {author}
  {\bibfnamefont {G.~F.}\ \bibnamefont {Counsell}}, \bibinfo {author}
  {\bibfnamefont {S.~J.}\ \bibnamefont {Fielding}}, \bibinfo {author}
  {\bibfnamefont {M.}~\bibnamefont {Gryaznevich}}, \bibinfo {author}
  {\bibfnamefont {R.}~\bibnamefont {Martin}}, \bibinfo {author} {\bibfnamefont
  {M.}~\bibnamefont {Price}}, \bibinfo {author} {\bibfnamefont
  {C.}~\bibnamefont {Roach}}, \bibinfo {author} {\bibfnamefont
  {V.}~\bibnamefont {Shevchenko}}, \bibinfo {author} {\bibfnamefont
  {M.}~\bibnamefont {Tournianski}}, \bibinfo {author} {\bibfnamefont
  {M.}~\bibnamefont {Valovic}}, \bibinfo {author} {\bibfnamefont {M.~J.}\
  \bibnamefont {Walsh}}, \ and\ \bibinfo {author} {\bibfnamefont {H.~R.}\
  \bibnamefont {Wilson}},\ }\href {\doibase 10.1063/1.1352595} {\bibfield
  {journal} {\bibinfo  {journal} {Physics of Plasmas}\ }\textbf {\bibinfo
  {volume} {8}},\ \bibinfo {pages} {2101} (\bibinfo {year} {2001})}\BibitemShut
  {NoStop}%
\bibitem [{\citenamefont {Riva}\ \emph {et~al.}(2019)\citenamefont {Riva},
  \citenamefont {Militello}, \citenamefont {Elmore}, \citenamefont {Omotani},
  \citenamefont {Dudson},\ and\ \citenamefont {Walkden}}]{Riva2019}%
  \BibitemOpen
  \bibfield  {author} {\bibinfo {author} {\bibfnamefont {F.}~\bibnamefont
  {Riva}}, \bibinfo {author} {\bibfnamefont {F.}~\bibnamefont {Militello}},
  \bibinfo {author} {\bibfnamefont {S.}~\bibnamefont {Elmore}}, \bibinfo
  {author} {\bibfnamefont {J.~T.}\ \bibnamefont {Omotani}}, \bibinfo {author}
  {\bibfnamefont {B.}~\bibnamefont {Dudson}}, \ and\ \bibinfo {author}
  {\bibfnamefont {N.~R.}\ \bibnamefont {Walkden}},\ }\href@noop {} {\bibfield
  {journal} {\bibinfo  {journal} {Plasma Physics and Controlled Fusion}\
  }\textbf {\bibinfo {volume} {61}} (\bibinfo {year} {2019})}\BibitemShut
  {NoStop}%
\bibitem [{\citenamefont {Dudson}\ \emph {et~al.}(2020)\citenamefont {Dudson},
  \citenamefont {Gracias}, \citenamefont {Jorge}, \citenamefont {Nielsen},
  \citenamefont {Olsen}, \citenamefont {Paolo}, \citenamefont {Silva},
  \citenamefont {Tamain}, \citenamefont {Ciralolo}, \citenamefont {Fedorczak},
  \citenamefont {Galassi}, \citenamefont {Madsen}, \citenamefont {Militello},
  \citenamefont {Nace}, \citenamefont {Rasmussen}, \citenamefont {Riva},\ and\
  \citenamefont {Serre}}]{Dudson2020}%
  \BibitemOpen
  \bibfield  {author} {\bibinfo {author} {\bibfnamefont {B.~D.}\ \bibnamefont
  {Dudson}}, \bibinfo {author} {\bibfnamefont {W.~A.}\ \bibnamefont {Gracias}},
  \bibinfo {author} {\bibfnamefont {R.}~\bibnamefont {Jorge}}, \bibinfo
  {author} {\bibfnamefont {A.~H.}\ \bibnamefont {Nielsen}}, \bibinfo {author}
  {\bibfnamefont {M.~B.}\ \bibnamefont {Olsen}}, \bibinfo {author}
  {\bibfnamefont {R.}~\bibnamefont {Paolo}}, \bibinfo {author} {\bibfnamefont
  {C.}~\bibnamefont {Silva}}, \bibinfo {author} {\bibfnamefont
  {P.}~\bibnamefont {Tamain}}, \bibinfo {author} {\bibfnamefont
  {G.}~\bibnamefont {Ciralolo}}, \bibinfo {author} {\bibfnamefont
  {N.}~\bibnamefont {Fedorczak}}, \bibinfo {author} {\bibfnamefont
  {D.}~\bibnamefont {Galassi}}, \bibinfo {author} {\bibfnamefont
  {J.}~\bibnamefont {Madsen}}, \bibinfo {author} {\bibfnamefont
  {F.}~\bibnamefont {Militello}}, \bibinfo {author} {\bibfnamefont
  {N.}~\bibnamefont {Nace}}, \bibinfo {author} {\bibfnamefont {J.~J.}\
  \bibnamefont {Rasmussen}}, \bibinfo {author} {\bibfnamefont {F.}~\bibnamefont
  {Riva}}, \ and\ \bibinfo {author} {\bibfnamefont {E.}~\bibnamefont {Serre}},\
  }\href {http://eprints.whiterose.ac.uk/165909/} {\bibfield  {journal}
  {\bibinfo  {journal} {Plasma Physics and Controlled Fusion}\ }\textbf
  {\bibinfo {volume} {In Press}} (\bibinfo {year} {2020})}\BibitemShut
  {NoStop}%
\bibitem [{\citenamefont {Militello}\ \emph {et~al.}(2013)\citenamefont
  {Militello}, \citenamefont {Tamain}, \citenamefont {Fundamenski},
  \citenamefont {Kirk}, \citenamefont {Naulin},\ and\ \citenamefont
  {Nielsen}}]{Militello2013}%
  \BibitemOpen
  \bibfield  {author} {\bibinfo {author} {\bibfnamefont {F.}~\bibnamefont
  {Militello}}, \bibinfo {author} {\bibfnamefont {P.}~\bibnamefont {Tamain}},
  \bibinfo {author} {\bibfnamefont {W.}~\bibnamefont {Fundamenski}}, \bibinfo
  {author} {\bibfnamefont {A.}~\bibnamefont {Kirk}}, \bibinfo {author}
  {\bibfnamefont {V.}~\bibnamefont {Naulin}}, \ and\ \bibinfo {author}
  {\bibfnamefont {A.~H.}\ \bibnamefont {Nielsen}},\ }\href {\doibase
  10.1088/0741-3335/55/2/025005} {\bibfield  {journal} {\bibinfo  {journal}
  {Plasma Physics and Controlled Fusion}\ }\textbf {\bibinfo {volume} {55}},\
  \bibinfo {pages} {25005} (\bibinfo {year} {2013})},\ \Eprint
  {http://arxiv.org/abs/arXiv:1305.5064v1} {arXiv:arXiv:1305.5064v1}
  \BibitemShut {NoStop}%
\bibitem [{\citenamefont {Garcia}\ \emph {et~al.}(2006)\citenamefont {Garcia},
  \citenamefont {Bian},\ and\ \citenamefont {Fundamenski}}]{Garcia2006}%
  \BibitemOpen
  \bibfield  {author} {\bibinfo {author} {\bibfnamefont {O.~E.}\ \bibnamefont
  {Garcia}}, \bibinfo {author} {\bibfnamefont {N.~H.}\ \bibnamefont {Bian}}, \
  and\ \bibinfo {author} {\bibfnamefont {W.}~\bibnamefont {Fundamenski}},\
  }\href {\doibase 10.1063/1.2336422} {\bibfield  {journal} {\bibinfo
  {journal} {Physics of Plasmas}\ }\textbf {\bibinfo {volume} {13}} (\bibinfo
  {year} {2006}),\ 10.1063/1.2336422}\BibitemShut {NoStop}%
\bibitem [{\citenamefont {Myra}\ \emph
  {et~al.}(2011{\natexlab{b}})\citenamefont {Myra}, \citenamefont {Russell},
  \citenamefont {D'Ippolito}, \citenamefont {Ahn}, \citenamefont {Maingi},
  \citenamefont {Maqueda}, \citenamefont {Lundberg}, \citenamefont {Stotler},
  \citenamefont {Zweben},\ and\ \citenamefont {Umansky}}]{Myra2011a}%
  \BibitemOpen
  \bibfield  {author} {\bibinfo {author} {\bibfnamefont {J.~R.}\ \bibnamefont
  {Myra}}, \bibinfo {author} {\bibfnamefont {D.~A.}\ \bibnamefont {Russell}},
  \bibinfo {author} {\bibfnamefont {D.~A.}\ \bibnamefont {D'Ippolito}},
  \bibinfo {author} {\bibfnamefont {J.~W.}\ \bibnamefont {Ahn}}, \bibinfo
  {author} {\bibfnamefont {R.}~\bibnamefont {Maingi}}, \bibinfo {author}
  {\bibfnamefont {R.~J.}\ \bibnamefont {Maqueda}}, \bibinfo {author}
  {\bibfnamefont {D.~P.}\ \bibnamefont {Lundberg}}, \bibinfo {author}
  {\bibfnamefont {D.~P.}\ \bibnamefont {Stotler}}, \bibinfo {author}
  {\bibfnamefont {S.~J.}\ \bibnamefont {Zweben}}, \ and\ \bibinfo {author}
  {\bibfnamefont {M.}~\bibnamefont {Umansky}},\ }in\ \href {\doibase
  10.1016/j.jnucmat.2010.10.030} {\emph {\bibinfo {booktitle} {Journal of
  Nuclear Materials}}},\ Vol.\ \bibinfo {volume} {415}\ (\bibinfo {year}
  {2011})\ pp.\ \bibinfo {pages} {S605--S608}\BibitemShut {NoStop}%
\bibitem [{\citenamefont {D'Ippolito}\ \emph {et~al.}(2011)\citenamefont
  {D'Ippolito}, \citenamefont {Myra},\ and\ \citenamefont
  {Zweben}}]{DIppolito2011}%
  \BibitemOpen
  \bibfield  {author} {\bibinfo {author} {\bibfnamefont {D.~A.}\ \bibnamefont
  {D'Ippolito}}, \bibinfo {author} {\bibfnamefont {J.~R.}\ \bibnamefont
  {Myra}}, \ and\ \bibinfo {author} {\bibfnamefont {S.~J.}\ \bibnamefont
  {Zweben}},\ }\href {\doibase 10.1063/1.3594609} {\bibfield  {journal}
  {\bibinfo  {journal} {Physics of Plasmas}\ }\textbf {\bibinfo {volume} {18}}
  (\bibinfo {year} {2011}),\ 10.1063/1.3594609}\BibitemShut {NoStop}%
\bibitem [{\citenamefont {Garcia}\ \emph {et~al.}(2004)\citenamefont {Garcia},
  \citenamefont {Naulin}, \citenamefont {Nielsen},\ and\ \citenamefont
  {Rasmussen}}]{Garcia2004}%
  \BibitemOpen
  \bibfield  {author} {\bibinfo {author} {\bibfnamefont {O.~E.}\ \bibnamefont
  {Garcia}}, \bibinfo {author} {\bibfnamefont {V.}~\bibnamefont {Naulin}},
  \bibinfo {author} {\bibfnamefont {A.~H.}\ \bibnamefont {Nielsen}}, \ and\
  \bibinfo {author} {\bibfnamefont {J.~J.}\ \bibnamefont {Rasmussen}},\ }\href
  {\doibase 10.1103/PhysRevLett.92.165003} {\bibfield  {journal} {\bibinfo
  {journal} {Physical Review Letters}\ }\textbf {\bibinfo {volume} {92}},\
  \bibinfo {pages} {165003} (\bibinfo {year} {2004})},\ \Eprint
  {http://arxiv.org/abs/0309020} {arXiv:0309020 [physics]} \BibitemShut
  {NoStop}%
\bibitem [{\citenamefont {Dudson}\ \emph {et~al.}(2009)\citenamefont {Dudson},
  \citenamefont {Umansky}, \citenamefont {Xu}, \citenamefont {Snyder},\ and\
  \citenamefont {Wilson}}]{Dudson2009}%
  \BibitemOpen
  \bibfield  {author} {\bibinfo {author} {\bibfnamefont {B.~D.}\ \bibnamefont
  {Dudson}}, \bibinfo {author} {\bibfnamefont {M.~V.}\ \bibnamefont {Umansky}},
  \bibinfo {author} {\bibfnamefont {X.~Q.}\ \bibnamefont {Xu}}, \bibinfo
  {author} {\bibfnamefont {P.~B.}\ \bibnamefont {Snyder}}, \ and\ \bibinfo
  {author} {\bibfnamefont {H.~R.}\ \bibnamefont {Wilson}},\ }\href {\doibase
  10.1016/j.cpc.2009.03.008} {\bibfield  {journal} {\bibinfo  {journal}
  {Computer Physics Communications}\ }\textbf {\bibinfo {volume} {180}},\
  \bibinfo {pages} {1467} (\bibinfo {year} {2009})},\ \Eprint
  {http://arxiv.org/abs/0810.5757} {arXiv:0810.5757} \BibitemShut {NoStop}%
\bibitem [{\citenamefont {Dudson}\ \emph {et~al.}(2014)\citenamefont {Dudson},
  \citenamefont {Allen}, \citenamefont {Breyiannis}, \citenamefont {Brugger},
  \citenamefont {Buchanan}, \citenamefont {Easy}, \citenamefont {Farley},
  \citenamefont {Joseph}, \citenamefont {Kim}, \citenamefont {McGann},
  \citenamefont {Omotani}, \citenamefont {Umansky}, \citenamefont {Walkden},
  \citenamefont {Xia},\ and\ \citenamefont {Xu}}]{Dudson2014}%
  \BibitemOpen
  \bibfield  {author} {\bibinfo {author} {\bibfnamefont {B.~D.}\ \bibnamefont
  {Dudson}}, \bibinfo {author} {\bibfnamefont {a.}~\bibnamefont {Allen}},
  \bibinfo {author} {\bibfnamefont {G.}~\bibnamefont {Breyiannis}}, \bibinfo
  {author} {\bibfnamefont {E.}~\bibnamefont {Brugger}}, \bibinfo {author}
  {\bibfnamefont {J.}~\bibnamefont {Buchanan}}, \bibinfo {author}
  {\bibfnamefont {L.}~\bibnamefont {Easy}}, \bibinfo {author} {\bibfnamefont
  {S.}~\bibnamefont {Farley}}, \bibinfo {author} {\bibfnamefont
  {I.}~\bibnamefont {Joseph}}, \bibinfo {author} {\bibfnamefont
  {M.}~\bibnamefont {Kim}}, \bibinfo {author} {\bibfnamefont {a.~D.}\
  \bibnamefont {McGann}}, \bibinfo {author} {\bibfnamefont {J.~T.}\
  \bibnamefont {Omotani}}, \bibinfo {author} {\bibfnamefont {M.~V.}\
  \bibnamefont {Umansky}}, \bibinfo {author} {\bibfnamefont {N.~R.}\
  \bibnamefont {Walkden}}, \bibinfo {author} {\bibfnamefont {T.}~\bibnamefont
  {Xia}}, \ and\ \bibinfo {author} {\bibfnamefont {X.~Q.}\ \bibnamefont {Xu}},\
  }\href {\doibase 10.1017/S0022377814000816} {\bibfield  {journal} {\bibinfo
  {journal} {Journal of Plasma Physics}\ }\textbf {\bibinfo {volume} {81}}
  (\bibinfo {year} {2014}),\ 10.1017/S0022377814000816},\ \Eprint
  {http://arxiv.org/abs/1405.7905} {arXiv:1405.7905} \BibitemShut {NoStop}%
\bibitem [{\citenamefont {Dudson}\ \emph {et~al.}(2016)\citenamefont {Dudson},
  \citenamefont {Madsen}, \citenamefont {Omotani}, \citenamefont {Hill},
  \citenamefont {Easy},\ and\ \citenamefont {L{\o}iten}}]{Dudson2016a}%
  \BibitemOpen
  \bibfield  {author} {\bibinfo {author} {\bibfnamefont {B.~D.}\ \bibnamefont
  {Dudson}}, \bibinfo {author} {\bibfnamefont {J.}~\bibnamefont {Madsen}},
  \bibinfo {author} {\bibfnamefont {J.}~\bibnamefont {Omotani}}, \bibinfo
  {author} {\bibfnamefont {P.}~\bibnamefont {Hill}}, \bibinfo {author}
  {\bibfnamefont {L.}~\bibnamefont {Easy}}, \ and\ \bibinfo {author}
  {\bibfnamefont {M.}~\bibnamefont {L{\o}iten}},\ }\href {\doibase
  10.1063/1.4953429} {\bibfield  {journal} {\bibinfo  {journal} {Physics of
  Plasmas}\ }\textbf {\bibinfo {volume} {23}} (\bibinfo {year} {2016}),\
  10.1063/1.4953429},\ \Eprint {http://arxiv.org/abs/1602.06747}
  {arXiv:1602.06747} \BibitemShut {NoStop}%
\bibitem [{\citenamefont {Walkden}\ \emph {et~al.}(2019)\citenamefont
  {Walkden}, \citenamefont {Riva}, \citenamefont {Dudson}, \citenamefont {Ham},
  \citenamefont {Militello}, \citenamefont {Moulton}, \citenamefont
  {Nicholas},\ and\ \citenamefont {Omotani}}]{Walkden2019}%
  \BibitemOpen
  \bibfield  {author} {\bibinfo {author} {\bibfnamefont {N.~R.}\ \bibnamefont
  {Walkden}}, \bibinfo {author} {\bibfnamefont {F.}~\bibnamefont {Riva}},
  \bibinfo {author} {\bibfnamefont {B.~D.}\ \bibnamefont {Dudson}}, \bibinfo
  {author} {\bibfnamefont {C.}~\bibnamefont {Ham}}, \bibinfo {author}
  {\bibfnamefont {F.}~\bibnamefont {Militello}}, \bibinfo {author}
  {\bibfnamefont {D.}~\bibnamefont {Moulton}}, \bibinfo {author} {\bibfnamefont
  {T.}~\bibnamefont {Nicholas}}, \ and\ \bibinfo {author} {\bibfnamefont
  {J.~T.}\ \bibnamefont {Omotani}},\ }\href {\doibase
  10.1016/j.nme.2018.12.005} {\bibfield  {journal} {\bibinfo  {journal}
  {Nuclear Materials and Energy}\ }\textbf {\bibinfo {volume} {18}},\ \bibinfo
  {pages} {111} (\bibinfo {year} {2019})}\BibitemShut {NoStop}%
\bibitem [{\citenamefont {Walkden}(2015)}]{Walkden2015}%
  \BibitemOpen
  \bibfield  {author} {\bibinfo {author} {\bibfnamefont {N.}~\bibnamefont
  {Walkden}},\ }\href@noop {} {\emph {\bibinfo {title} {{Derivation of Electron
  Temperature Equation}}}},\ \bibinfo {type} {Tech. Rep.}\ \bibinfo {number}
  {January}\ (\bibinfo {year} {2015})\BibitemShut {NoStop}%
\bibitem [{\citenamefont {Easy}\ \emph {et~al.}(2016)\citenamefont {Easy},
  \citenamefont {Militello}, \citenamefont {Omotani}, \citenamefont {Walkden},\
  and\ \citenamefont {Dudson}}]{Easy2016a}%
  \BibitemOpen
  \bibfield  {author} {\bibinfo {author} {\bibfnamefont {L.}~\bibnamefont
  {Easy}}, \bibinfo {author} {\bibfnamefont {F.}~\bibnamefont {Militello}},
  \bibinfo {author} {\bibfnamefont {J.}~\bibnamefont {Omotani}}, \bibinfo
  {author} {\bibfnamefont {N.~R.}\ \bibnamefont {Walkden}}, \ and\ \bibinfo
  {author} {\bibfnamefont {B.}~\bibnamefont {Dudson}},\ }\href {\doibase
  10.1063/1.4940330} {\bibfield  {journal} {\bibinfo  {journal} {Physics of
  Plasmas}\ }\textbf {\bibinfo {volume} {23}} (\bibinfo {year} {2016}),\
  10.1063/1.4940330},\ \Eprint {http://arxiv.org/abs/1508.04085}
  {arXiv:1508.04085} \BibitemShut {NoStop}%
\bibitem [{\citenamefont {Walkden}\ \emph {et~al.}(2016)\citenamefont
  {Walkden}, \citenamefont {Easy}, \citenamefont {Militello},\ and\
  \citenamefont {Omotani}}]{Walkden2016}%
  \BibitemOpen
  \bibfield  {author} {\bibinfo {author} {\bibfnamefont {N.~R.}\ \bibnamefont
  {Walkden}}, \bibinfo {author} {\bibfnamefont {L.}~\bibnamefont {Easy}},
  \bibinfo {author} {\bibfnamefont {F.}~\bibnamefont {Militello}}, \ and\
  \bibinfo {author} {\bibfnamefont {J.~T.}\ \bibnamefont {Omotani}},\ }\href
  {\doibase 10.1088/0741-3335/58/11/115010} {\bibfield  {journal} {\bibinfo
  {journal} {Plasma Physics and Controlled Fusion}\ }\textbf {\bibinfo {volume}
  {58}} (\bibinfo {year} {2016}),\ 10.1088/0741-3335/58/11/115010},\ \Eprint
  {http://arxiv.org/abs/1606.00582} {arXiv:1606.00582} \BibitemShut {NoStop}%
\bibitem [{\citenamefont {Militello}\ \emph {et~al.}(2017)\citenamefont
  {Militello}, \citenamefont {Dudson}, \citenamefont {Easy}, \citenamefont
  {Kirk},\ and\ \citenamefont {Naylor}}]{Militello2017}%
  \BibitemOpen
  \bibfield  {author} {\bibinfo {author} {\bibfnamefont {F.}~\bibnamefont
  {Militello}}, \bibinfo {author} {\bibfnamefont {B.}~\bibnamefont {Dudson}},
  \bibinfo {author} {\bibfnamefont {L.}~\bibnamefont {Easy}}, \bibinfo {author}
  {\bibfnamefont {A.}~\bibnamefont {Kirk}}, \ and\ \bibinfo {author}
  {\bibfnamefont {P.}~\bibnamefont {Naylor}},\ }\href {\doibase
  10.1088/1361-6587/aa9252} {\bibfield  {journal} {\bibinfo  {journal} {Plasma
  Physics and Controlled Fusion}\ }\textbf {\bibinfo {volume} {59}} (\bibinfo
  {year} {2017}),\ 10.1088/1361-6587/aa9252}\BibitemShut {NoStop}%
\bibitem [{\citenamefont {Hoare}\ \emph {et~al.}(2019)\citenamefont {Hoare},
  \citenamefont {Militello}, \citenamefont {Omotani}, \citenamefont {Riva},\
  and\ \citenamefont {Newton}}]{Hoare2019}%
  \BibitemOpen
  \bibfield  {author} {\bibinfo {author} {\bibfnamefont {D.}~\bibnamefont
  {Hoare}}, \bibinfo {author} {\bibfnamefont {F.}~\bibnamefont {Militello}},
  \bibinfo {author} {\bibfnamefont {J.~T.}\ \bibnamefont {Omotani}}, \bibinfo
  {author} {\bibfnamefont {F.}~\bibnamefont {Riva}}, \ and\ \bibinfo {author}
  {\bibfnamefont {S.}~\bibnamefont {Newton}},\ }\href@noop {} {\bibfield
  {journal} {\bibinfo  {journal} {Plasma Physics and Controlled Fusion}\
  }\textbf {\bibinfo {volume} {61}} (\bibinfo {year} {2019})}\BibitemShut
  {NoStop}%
\bibitem [{\citenamefont {Schw{\"{o}}rer}\ \emph {et~al.}(2017)\citenamefont
  {Schw{\"{o}}rer}, \citenamefont {Walkden}, \citenamefont {Leggate},
  \citenamefont {Dudson}, \citenamefont {Militello}, \citenamefont {Downes},\
  and\ \citenamefont {Turner}}]{Schworer2017}%
  \BibitemOpen
  \bibfield  {author} {\bibinfo {author} {\bibfnamefont {D.}~\bibnamefont
  {Schw{\"{o}}rer}}, \bibinfo {author} {\bibfnamefont {N.~R.}\ \bibnamefont
  {Walkden}}, \bibinfo {author} {\bibfnamefont {H.}~\bibnamefont {Leggate}},
  \bibinfo {author} {\bibfnamefont {B.~D.}\ \bibnamefont {Dudson}}, \bibinfo
  {author} {\bibfnamefont {F.}~\bibnamefont {Militello}}, \bibinfo {author}
  {\bibfnamefont {T.}~\bibnamefont {Downes}}, \ and\ \bibinfo {author}
  {\bibfnamefont {M.~M.}\ \bibnamefont {Turner}},\ }\href {\doibase
  10.1016/j.nme.2017.02.016} {\bibfield  {journal} {\bibinfo  {journal}
  {Nuclear Materials and Energy}\ }\textbf {\bibinfo {volume} {12}},\ \bibinfo
  {pages} {825} (\bibinfo {year} {2017})}\BibitemShut {NoStop}%
\bibitem [{\citenamefont {Schworer}\ \emph {et~al.}(2018)\citenamefont
  {Schworer}, \citenamefont {Walkden}, \citenamefont {Leggate}, \citenamefont
  {Dudson}, \citenamefont {Militello}, \citenamefont {Downes},\ and\
  \citenamefont {Turner}}]{Schworer2018}%
  \BibitemOpen
  \bibfield  {author} {\bibinfo {author} {\bibfnamefont {D.}~\bibnamefont
  {Schworer}}, \bibinfo {author} {\bibfnamefont {N.~R.}\ \bibnamefont
  {Walkden}}, \bibinfo {author} {\bibfnamefont {H.}~\bibnamefont {Leggate}},
  \bibinfo {author} {\bibfnamefont {B.~D.}\ \bibnamefont {Dudson}}, \bibinfo
  {author} {\bibfnamefont {F.}~\bibnamefont {Militello}}, \bibinfo {author}
  {\bibfnamefont {T.}~\bibnamefont {Downes}}, \ and\ \bibinfo {author}
  {\bibfnamefont {M.~M.}\ \bibnamefont {Turner}},\ }\href@noop {} {\bibfield
  {journal} {\bibinfo  {journal} {Plasma Physics and Controlled Fusion}\
  }\textbf {\bibinfo {volume} {61}} (\bibinfo {year} {2018})}\BibitemShut
  {NoStop}%
\bibitem [{\citenamefont {Braginskii}(1965)}]{BraginskiiS1965}%
  \BibitemOpen
  \bibfield  {author} {\bibinfo {author} {\bibfnamefont {S.~I.}\ \bibnamefont
  {Braginskii}},\ }\href@noop {} {\bibfield  {journal} {\bibinfo  {journal}
  {Reviews of Plasma Physics}\ } (\bibinfo {year} {1965})}\BibitemShut
  {NoStop}%
\bibitem [{\citenamefont {Stangeby}(2000)}]{Stangeby2000}%
  \BibitemOpen
  \bibfield  {author} {\bibinfo {author} {\bibfnamefont {P.~C.}\ \bibnamefont
  {Stangeby}},\ }\href {\doibase 10.1088/0741-3335/43/2/702} {\emph {\bibinfo
  {title} {Plasma Physics and Controlled Fusion}}},\ Vol.~\bibinfo {volume}
  {43}\ (\bibinfo {year} {2000})\ pp.\ \bibinfo {pages} {223--224}\BibitemShut
  {NoStop}%
\bibitem [{\citenamefont {Fundamenski}\ \emph {et~al.}(2007)\citenamefont
  {Fundamenski}, \citenamefont {Garcia}, \citenamefont {Naulin}, \citenamefont
  {Pitts}, \citenamefont {Nielsen}, \citenamefont {{Juul Rasmussen}},
  \citenamefont {Horacek},\ and\ \citenamefont {Graves}}]{Fundamenski2007}%
  \BibitemOpen
  \bibfield  {author} {\bibinfo {author} {\bibfnamefont {W.}~\bibnamefont
  {Fundamenski}}, \bibinfo {author} {\bibfnamefont {O.~E.}\ \bibnamefont
  {Garcia}}, \bibinfo {author} {\bibfnamefont {V.}~\bibnamefont {Naulin}},
  \bibinfo {author} {\bibfnamefont {R.~A.}\ \bibnamefont {Pitts}}, \bibinfo
  {author} {\bibfnamefont {A.~H.}\ \bibnamefont {Nielsen}}, \bibinfo {author}
  {\bibfnamefont {J.}~\bibnamefont {{Juul Rasmussen}}}, \bibinfo {author}
  {\bibfnamefont {J.}~\bibnamefont {Horacek}}, \ and\ \bibinfo {author}
  {\bibfnamefont {J.~P.}\ \bibnamefont {Graves}},\ }\href {\doibase
  10.1088/0029-5515/47/5/006} {\bibfield  {journal} {\bibinfo  {journal}
  {Nuclear Fusion}\ }\textbf {\bibinfo {volume} {47}},\ \bibinfo {pages} {417}
  (\bibinfo {year} {2007})}\BibitemShut {NoStop}%
\bibitem [{\citenamefont {Antar}\ \emph {et~al.}(2003)\citenamefont {Antar},
  \citenamefont {Counsell}, \citenamefont {Yu}, \citenamefont {Labombard},\
  and\ \citenamefont {Devynck}}]{Antar2003}%
  \BibitemOpen
  \bibfield  {author} {\bibinfo {author} {\bibfnamefont {G.~Y.}\ \bibnamefont
  {Antar}}, \bibinfo {author} {\bibfnamefont {G.~F.}\ \bibnamefont {Counsell}},
  \bibinfo {author} {\bibfnamefont {Y.}~\bibnamefont {Yu}}, \bibinfo {author}
  {\bibfnamefont {B.}~\bibnamefont {Labombard}}, \ and\ \bibinfo {author}
  {\bibfnamefont {P.}~\bibnamefont {Devynck}},\ }\href {\doibase
  10.1002/ctpp.200410031} {\bibfield  {journal} {\bibinfo  {journal} {Physics
  of Plasmas}\ }\textbf {\bibinfo {volume} {10}},\ \bibinfo {pages} {419}
  (\bibinfo {year} {2003})}\BibitemShut {NoStop}%
\bibitem [{\citenamefont {Militello}\ and\ \citenamefont
  {Fundamenski}(2011)}]{Militello2011}%
  \BibitemOpen
  \bibfield  {author} {\bibinfo {author} {\bibfnamefont {F.}~\bibnamefont
  {Militello}}\ and\ \bibinfo {author} {\bibfnamefont {W.}~\bibnamefont
  {Fundamenski}},\ }\href {\doibase 10.1088/0741-3335/53/9/095002} {\bibfield
  {journal} {\bibinfo  {journal} {Plasma Physics and Controlled Fusion}\
  }\textbf {\bibinfo {volume} {53}},\ \bibinfo {pages} {095002} (\bibinfo
  {year} {2011})}\BibitemShut {NoStop}%
\bibitem [{\citenamefont {Dudson}\ \emph {et~al.}(2019)\citenamefont {Dudson},
  \citenamefont {Allen}, \citenamefont {Body}, \citenamefont {Chapman},
  \citenamefont {Lau}, \citenamefont {Townley}, \citenamefont {Moulton},
  \citenamefont {Harrison},\ and\ \citenamefont {Lipschultz}}]{Dudson2019a}%
  \BibitemOpen
  \bibfield  {author} {\bibinfo {author} {\bibfnamefont {B.~D.}\ \bibnamefont
  {Dudson}}, \bibinfo {author} {\bibfnamefont {J.}~\bibnamefont {Allen}},
  \bibinfo {author} {\bibfnamefont {T.}~\bibnamefont {Body}}, \bibinfo {author}
  {\bibfnamefont {B.}~\bibnamefont {Chapman}}, \bibinfo {author} {\bibfnamefont
  {C.}~\bibnamefont {Lau}}, \bibinfo {author} {\bibfnamefont {L.}~\bibnamefont
  {Townley}}, \bibinfo {author} {\bibfnamefont {D.}~\bibnamefont {Moulton}},
  \bibinfo {author} {\bibfnamefont {J.}~\bibnamefont {Harrison}}, \ and\
  \bibinfo {author} {\bibfnamefont {B.}~\bibnamefont {Lipschultz}},\
  }\href@noop {} {\bibfield  {journal} {\bibinfo  {journal} {Plasma Physics and
  Controlled Fusion}\ }\textbf {\bibinfo {volume} {61}} (\bibinfo {year}
  {2019})}\BibitemShut {NoStop}%
\bibitem [{\citenamefont {Ghendrih}\ \emph {et~al.}(2011)\citenamefont
  {Ghendrih}, \citenamefont {Bodi}, \citenamefont {Bufferand}, \citenamefont
  {Chiavassa}, \citenamefont {Ciraolo}, \citenamefont {Fedorczak},
  \citenamefont {Isoardi}, \citenamefont {Paredes}, \citenamefont {Sarazin},
  \citenamefont {Serre}, \citenamefont {Schwander},\ and\ \citenamefont
  {Tamain}}]{Ghendrih2011}%
  \BibitemOpen
  \bibfield  {author} {\bibinfo {author} {\bibfnamefont {P.}~\bibnamefont
  {Ghendrih}}, \bibinfo {author} {\bibfnamefont {K.}~\bibnamefont {Bodi}},
  \bibinfo {author} {\bibfnamefont {H.}~\bibnamefont {Bufferand}}, \bibinfo
  {author} {\bibfnamefont {G.}~\bibnamefont {Chiavassa}}, \bibinfo {author}
  {\bibfnamefont {G.}~\bibnamefont {Ciraolo}}, \bibinfo {author} {\bibfnamefont
  {N.}~\bibnamefont {Fedorczak}}, \bibinfo {author} {\bibfnamefont
  {L.}~\bibnamefont {Isoardi}}, \bibinfo {author} {\bibfnamefont
  {A.}~\bibnamefont {Paredes}}, \bibinfo {author} {\bibfnamefont
  {Y.}~\bibnamefont {Sarazin}}, \bibinfo {author} {\bibfnamefont
  {E.}~\bibnamefont {Serre}}, \bibinfo {author} {\bibfnamefont
  {F.}~\bibnamefont {Schwander}}, \ and\ \bibinfo {author} {\bibfnamefont
  {P.}~\bibnamefont {Tamain}},\ }\href {\doibase 10.1088/0741-3335/53/5/054019}
  {\bibfield  {journal} {\bibinfo  {journal} {Plasma Physics and Controlled
  Fusion}\ }\textbf {\bibinfo {volume} {53}} (\bibinfo {year} {2011}),\
  10.1088/0741-3335/53/5/054019}\BibitemShut {NoStop}%
\end{thebibliography}%

\end{document}